\documentclass[twocolumn, superscriptaddress, amsmath, amssymb, aps, prx]{revtex4-2}

\usepackage{graphicx}% Include figure files
\usepackage{dcolumn}% Align table columns on decimal point
\usepackage{bm}% bold math
\usepackage{xcolor}
\usepackage{mathtools}
\usepackage{physics}
\usepackage[normalem]{ulem}
\usepackage{soul}
\usepackage{braket}
\usepackage{media9}

\usepackage{hyperref}
\hypersetup{colorlinks=true, citecolor=blue, linkcolor=blue, urlcolor=blue}

\begin{document}

\title{Novel Topology and Manipulation of Scattering Singularities in Complex non-Hermitian Systems}

\author{Jared Erb}
 \email{jmerb@umd.edu}
 \affiliation{Maryland Quantum Materials Center, Department of Physics, University of Maryland, College Park, MD, 20742, USA}

\author{Nadav Shaibe}
 \affiliation{Maryland Quantum Materials Center, Department of Physics, University of Maryland, College Park, MD, 20742, USA}

\author{Robert Calvo}
 \affiliation{Wave Transport in Complex Systems Lab, Physics Department, Wesleyan University, Middletown, CT, 06459, USA}

\author{Daniel P. Lathrop}
 \affiliation{Institute for Research in Electronics and Applied Physics, Department
of Physics, University of Maryland, College Park, MD, 20742, USA}

\author{Thomas M. Antonsen Jr.}
 \affiliation{Institute for Research in Electronics and Applied Physics, Department
of Physics, University of Maryland, College Park, MD, 20742, USA}

\author{Tsampikos Kottos}
 \affiliation{Wave Transport in Complex Systems Lab, Physics Department, Wesleyan University, Middletown, CT, 06459, USA}

\author{Steven M. Anlage}
 \affiliation{Maryland Quantum Materials Center, Department of Physics, University of Maryland, College Park, MD, 20742, USA}

\date{\today}

\begin{abstract}

The control of wave scattering in complex non-Hermitian settings is an exciting subject -- often challenging the creativity of researchers and stimulating the imagination of the public. Successful outcomes include invisibility cloaks, wavefront shaping protocols, active metasurface development, and more. At their core, these achievements rely on our ability to engineer the resonant spectrum of the underlying physical structures which is conventionally accomplished by carefully imposing geometrical and/or dynamical symmetries. In contrast, by taking active control over the boundary conditions in complex scattering environments which lack artificially-imposed geometric symmetries, we demonstrate via microwave experiments the ability to manipulate the spectrum of the scattering operator. This active control empowers the creation, destruction and repositioning of exceptional point degeneracies (EPD's) in a two-dimensional (2D) parameter space. The presence of EPD's signifies a coalescence of the scattering eigenmodes, which dramatically affects transport. The scattering EPD's are partitioned in domains characterized by a binary charge, as well as an integer winding number, are topologically stable in the two-dimensional parameter space, and obey winding number-conservation laws upon interactions with each other, even in cases where Lorentz reciprocity is violated; in this case the topological domains are destroyed. Ramifications of this understanding is the proposition for a unique input-magnitude and phase-insensitive 50:50 in-phase/quadrature (I/Q) power splitter. Our study establishes an important step towards complete control of scattering processes in complex non-Hermitian settings.\\

\end{abstract}


\maketitle

\section{Introduction}\label{sec1}

Much of modern physics is based on the study of Hermitian quantum and classical systems in which the effects of loss or gain are either excluded or ignored. This ensures that the eigenvalues of the Hamiltonian that describes such systems are real and the eigenvectors are orthogonal. However, by including interactions of the system with the environment, the resulting effective Hamiltonian is no longer Hermitian. This non-hermiticity generally renders the eigenvalues complex, and the eigenvectors are no longer orthogonal \cite{Bender98}. Nevertheless, it has been realized that non-Hermitian Hamiltonian systems offer qualitatively new phenomena and features that are not found in the Hermitian case. These include the possibility of Coherent Perfect Absorption (CPA), exceptional point degeneracies (EPD) of the frequency eigenmodes, and others \cite{Chong10,Feng17,Baranov17,El-Ganainy18,Alu19,Özdemir19,Ashida20}. 

Undoubtedly, among the various non-Hermitian peculiarities, \textit{resonant} EPD's and the physical phenomena emanating from their presence is the tip of the spear of many of the new developments, from the fields of optics \cite{PhysRevLett.106.213901,PhysRevLett.112.203901,Hodaei2017,Chen2017}, microwave cavities \cite{PhysRevLett.86.787,PhysRevE.69.056216,PhysRevLett.106.150403}, exciton-polaritons \cite{Gao2015}, acoustics \cite{Shi2016}, waveguides \cite{Doppler2016}, electrical circuits \cite{Stehmann_2004,Suntharalingam2023}, etc. An EPD of order $P$ (EPD-P) occurs when $P\geq2$ eigenvalues and the associated eigenvectors of a non-Hermitian operator coalesce at a specific point in the parameter space of the system \cite{Berry2004,Ding2022}. A topological characterization of these singularities has been intensively studied in the framework of non-Hermitian operators \cite{PhysRevE.69.056216,Ding2022}. The eigenbasis collapse has dramatic consequences, the most prominent being the enhancement of wave-matter interaction in the proximity of EPD's. In their proximity, the eigenfrequency Riemann surfaces (RS) self-intersect, exhibiting a complex non-trivial topology which is typically characterized by a fractional perturbation expansion of the eigenvalue detuning $\delta\lambda \equiv \lambda - \lambda_{EPD} \sim \sum_kc_k \varepsilon^{k/P}$, where $\varepsilon$ is a small variation of one of the control parameters of the system from its EPD value. The extreme sensitivity to small parameter variations has recently attracted a lot of interest in using EPD's as sensors \cite{PhysRevLett.112.203901,Hodaei2017,Chen2017,Kononchuk22}. Another manifestation of the anomalous topological features of the RS near an EPD emerges when they are encircled by means of a variation of the parameters of the system. If the parametric variation of the Hamiltonian around an EPD occurs quasistatically, the instantaneous eigenstates transform into each other at the end of the cycle with only one of them acquiring a geometric phase \cite{PhysRevLett.86.787,PhysRevE.69.056216}. If, however, the evolution around the EPD is dynamical (but still slow) then only one state dominates the output and what determines this preferred eigenstate is the sense of rotation in the parameter space \cite{Doppler2016,Zhong2018,Khajavikhan22,Nasari22,Guria2024}. These two examples are representative of the novel transport features demonstrated by non-Hermitian systems when their resonances (associated with the spectrum of an effective Hamiltonian) are parametrically varied in the proximity of an EPD. 

In contrast with the mainstream literature on non-Hermitian singularities which discusses degeneracies in the resonance spectrum, here, we analyze a new type of EPD associated with the spectrum of sub (or super)-unitary operators (SU-EPD's). The most typical physical example of such an operator is the scattering matrix $S$. In this paper, we specifically consider EPD's in the spectrum of such class of super/sub-unitary operators, and unveil several new features that are not present in resonant EPD's occurring at the frequency spectrum of non-Hermitian Hamiltonians. 

The motivation to investigate the spectral singularities of scattering matrices is twofold: (a) the $S$-matrix is widely accessible in many experimental contexts, including acoustics, particle physics, quantum transport, non-destructive testing, and electromagnetic waves while (b) it contains a tremendous wealth of information, much of which has heretofore been considered inaccessible or un-interpretable, and therefore widely under-appreciated. 

The $M\times M-$dimensional scattering matrix $S$ maps an incoming set of waves injected to the scattering target via $M$ channels onto a set of outgoing waves $\ket{\psi_{out}} = S \ket{\psi_{in}}$, where the complex components of the $M$-dimensional vectors $\ket{\psi_{in/out}}$ denote the incoming/outgoing wave amplitudes of a monochromatic field in the channel representation. Since the scattering matrix is related to the resolvent of the effective Hamiltonian that describes the underlying system, it is expected to encode its spectral peculiarities -- and particularly the intriguing topological features of the EPD's. In this framework, the EPD's are associated with the eigenvalues of the sub(super)-unitary operator while the (injected) frequency acts as a new control-parameter. As such, it enables strong (global) perturbations to the system, resulting in the creation, manipulation, and annihilation of many scattering singularities. 

By utilizing, as additional parameters, strong electronically-tunable boundary perturbations, we find that the super/sub-unitary operators can be readily manipulated to show an abundance of EPD's, rather than individual examples normally obtained under highly symmetric conditions. These SU-EPD's evolve systematically as we perturb the system, have robust topological properties, and are pairwise created and annihilated in great abundance. We topologically characterize (through their binary charge and winding number) these SU-EPD's and identify rules that govern their creation and annihilation.

Based on the topological characterization of SU-EPD's we can partition the parameter space with domains that host SU-EPD's of specific charge. These topological domains are separated from one-another by orthogonality ``walls" where the scattering eigenvectors form a complete orthogonal basis. We analyze the resilience of these topological invariants in case of symmetry violations and show that a non-zero magnetic field can destroy the topological compartmentalization of the parameter space. Nevertheless, a new set of topological constraints is established on the evolution of the EPD's in this case.

The use of controllable metasurfaces embedded inside a scattering electromagnetic environment to ``pre-distort” a signal to achieve a certain goal (e.g. focusing light beyond or inside a strongly scattering medium, etc.) provides a realistic framework for the implementation of the above rules. An immediate application that emerges from the capability to manipulate the SU-EPD's using the above rules is an input-magnitude/phase-insensitive 50:50 I/Q-output power splitter, occurring for a special class of EPD's that are zeros of the scattering matrix. The results are general in the sense that they apply to all types of waves (scalar: quantum, acoustic, flexural, and vector: electromagnetic, etc.), span waves propagating in different dimensions (see examples in Fig. \ref{Schematics}), and media with or without Lorentz reciprocal properties. 

This paper is structured as follows.  Section \ref{Sec:SandEPD} describes the mathematical properties of scattering matrix exceptional points.  Section \ref{Sec:EandT} presents the data and models that reveal the ubiquity, general properties, and dynamical evolution of EPD's.  Section \ref{Sec:Robust5050} presents data for a novel scattering singularity application enabled by our understanding of EPD's, while Section \ref{Sec:Diss} discusses the results and concludes the paper.  Appendices \ref{sec.Limitations}-\ref{sec.Verification} document the details and additional data supporting the claims made in the text.

\section{Scattering matrix and Exceptional Points}
\label{Sec:SandEPD}
We consider complex multi-modal physical settings consisting of a scattering domain that is interrogated by $M=2$ channels, see Fig. \ref{Schematics}. The scattering properties of these systems are described by a $2\times 2$ scattering matrix $S$ whose complex matrix elements are, in general, irregular functions of excitation frequency. The scattering matrix elements are also sensitive to boundary perturbations and changes in uniform and lumped attenuation as well as the coupling of the system with the channels. For a general $2\times 2$ scattering matrix 
$S = \left(\begin{smallmatrix}
  S_{11} & S_{12}\\
  S_{21} & S_{22}
\end{smallmatrix}\right)$, the eigenvalues and eigenvectors take the form:

\begin{equation}
\lambda_S^{1,2} = \frac{S_{11}+S_{22}}{2} \pm \frac{1}{2}\sqrt{4 S_{12} S_{21} \left( 1 + M_{S}^2 \right)}
\label{eval}
\end{equation}

\begin{equation}
 \ket{R_{1,2}} =
\begin{pmatrix}
\frac{1}{S_{21}}\left(\lambda_S^{1,2} - S_{22}\right)\\
1
\end{pmatrix},
\label{R1}
\end{equation}
where
\begin{equation}
M_{S} \equiv \frac{S_{11} - S_{22}}{2 \sqrt{S_{12} S_{21}}}
\label{eqn:MSDef}
\end{equation}
is a complex scalar function of the system parameters and of incident frequency (some limitations of these expressions are discussed in Appendix \ref{sec.Limitations}). In the case that the $S$-matrix is sub/super-unitary the eigenvectors form a bi-orthogonal basis with the right eigenvectors $\{\ket{R_i}\}$ being defined by the relation $S \ket{R_i} = \lambda_i \ket{R_i}$. Similarly, the left eigenvectors are evaluated by the relation $\bra{L_i} S = \lambda_i \bra{L_i}$. In this case, we will be using the following orthonormality conventions:
\begin{gather} 
\left\langle R_i \middle| R_i \right\rangle = \left\langle L_i \middle| L_i \right\rangle = 1, \label{Norm_Biorth} 
\:\: \left\langle L_i \middle| R_j \right\rangle_{i \neq j} = 0,
\end{gather}
where the first equation is the normalization condition for the eigenvectors and the second equation is the biorthogonality condition of the left and right eigenvectors \cite{Ashida20}. 

In fact, there are points in the parameter space of the system where the eigenbasis collapses and the spectrum develops SU-EPD's. At these points, both eigenvalues and their associated eigenvectors coalesce resulting in a catastrophic event where the eigenbasis collapses. A measure of the proximity to these SU-EPD's is the eigenvector coalescence $|C|$ which is defined as

\begin{equation} \label{eqn:CDef}
\begin{split}
|C| & = \frac{1}{\dbinom{M}{2}} \sum\limits_{\substack{i=1,j=2 \\ i<j}}^M \frac{\lvert \langle R_i \vert R_j \rangle \rvert}{\lvert \lvert R_i\rvert \rvert \: \lvert \lvert R_j\rvert \rvert} \\
 & = \frac{1}{\dbinom{M}{2}} \sum\limits_{\substack{i=1,j=2 \\ i<j}}^M \frac{\lvert \langle L_i \vert L_j \rangle \rvert}{\lvert \lvert L_i\rvert \rvert \: \lvert \lvert L_j\rvert \rvert},
\end{split}
\end{equation}

where $\left\langle u \middle| v \right\rangle = u^{\dag} v$ is the inner product of the two eigenvectors, $||u|| = \sqrt{\left\langle u \middle| u \right\rangle}$ is the norm of the eigenvector, and $M$ is the number of channels. With these conventions, the coalescence $|C|$ is a real number between $0$ and $1$, with $|C|=1$ corresponding to full coalescence (degeneracy of the eigenvectors) and $|C|=0$ representing full orthogonality of the eigenvectors, the latter being significant for the topological characterization of the partitioning of the parameter space.

\subsection{Scattering EPD's and their topological characterization}

From Eqs. (\ref{eval},\ref{R1},\ref{eqn:MSDef}) we find that both the eigenvectors and eigenvalues become degenerate when $M_{S}=\pm i$. The degenerate eigenvectors take the form 
\begin{equation}
\ket{R_{\pm i}} = \frac{1}{\sqrt{1+\left| \frac{S_{12}}{S_{21}}\right|}}\begin{pmatrix} \frac{\pm i \sqrt{S_{12}S_{21}}}{S_{21}}\\ 1 \end{pmatrix}
\label{DV}
\end{equation}
indicating a collapse of the eigenbasis. The collapsed eigenbasis can be completed by introducing a Jordan vector $\ket{J}$ . The construction utilizes the Jordan chain scheme leading to the following expression for the Jordan vector \cite{Bronson69,Seyranian03}
\begin{equation}
S \ket{J_{\pm i}} = \lambda_{s}^{\pm i} \ket{J_{\pm i}} + \ket{R_{\pm i}}, 
\label{JV}
\end{equation}
where $\ket{R_{\pm i}}$ is the degenerate vector (see Eq. (\ref{DV})), and $\lambda_S^{\pm i}=\frac{S_{11}+S_{22}}{2}$ (see Eq. (\ref{eval})). For more information on the Jordan vector, see Appendix \ref{sec.Jordan}.

The \textit{orthogonality} of the eigenvectors is restored when the following \textit{two} conditions are satisfied simultaneously: ${\cal I}m(M_{S}) = 0$ \textit{and} $\left| \frac{S_{12}}{S_{21}} \right| = 1$ (see Appendix \ref{sec.Proof}). The violation of the second constraint enforces the violation of the reciprocity condition; thus guaranteeing the existence of the bi-orthogonal basis. Furthermore, the simultaneous satisfaction of the two constraints implies that the eigenvector orthogonality occurs in general at discrete points in a two-dimensional parameter space. 
%, as the coalescence $|C|$ is precisely zero when $M_{S}$ is purely real and $\left| \frac{S_{12}}{S_{21}} \right| = 1$ 

One can further distinguish the SU-EPD's $M_S=\pm i$ of opposite charge by realizing that they are independent because they correspond to phase-singularities (zeros) of different complex scalar functions. To this end, we rewrite the EPD condition for positive $(M_S=+i)$ and negative $(M_S=-i)$ charge as a vortex-condition for the functions $S_{\pm i}$:
\begin{gather} 
S_{+i} \equiv S_{11} - S_{22} - 2i\sqrt{S_{12}S_{21}} = \left| S_{+i} \right| e^{i\theta^+} \label{Splus}\\
S_{-i} \equiv S_{11} - S_{22} + 2i\sqrt{S_{12}S_{21}} = \left| S_{-i} \right| e^{i\theta^-}. \label{Sminus}
\end{gather}
The topological characterization of the associated vortices that occur at points of the parameter space where the scattering EPD condition is satisfied is done by determining the winding number $n$:

\begin{equation}
n = \frac{1}{2 \pi} \oint\limits_{C} \nabla \theta \,\mathrm{d}s,
\label{eqn:winding_num}
\end{equation}
where $C$ is a counterclockwise loop enclosing the singularity of the phase field $\theta$ of the complex scalar function of interest. A winding number $n = \pm 1$ is topologically stable, while winding numbers of $|n| \ge 2$ are topologically unstable and generally split into $|n|$ vortices of winding number $sgn(n)$ \cite{Neu90}. The winding number of each scattering singularity found in our numerical and experimental data is only $\pm 1$. We also find that when scattering singularities are created or annihilated, they always do so as pairs, and the elements of these pairs have opposite winding numbers. Therefore the overall winding number of the system is conserved.

Finally, we point out that every EPD in a two-dimensional parameter space is generically connected to exactly one other EPD by a finite curve. There is a family of finite curves that connect EPD's as long as their endpoints are $M_S = \pm i$, but without loss of generality we choose the simplest finite curve defined by the conditions ${\cal R}e(M_{S}) = 0$ and $-1 \le {\cal I}m(M_{S}) \le 1$. EPD's connected this way have a fixed relative charge/winding number relation. The paired EPD's either have the same charge and different winding number or they have opposite charge and the same winding number (see Appendix \ref{sec.Phase_Winding} for further details).

\subsection{Special case of reciprocal scattering processes}

The non-Hermitian systems we consider here break time-reversal invariance due to the existence of gain and/or loss.  An additional symmetry is that of Lorentz reciprocity in transport through the system. The case of reciprocal scattering is of particular interest since many of the above complex functions used for the topological characterization of EPD's take a simple form. For a $2\times 2$ reciprocal scattering matrix ($S_{12} = S_{21}$), the equations for the eigenvalues and eigenvectors simplify (see Appendix \ref{sec.Simplification}) while $M_{S}$ becomes

\begin{equation}
M_S^R \equiv \frac{S_{11} - S_{22}}{2 S_{21}}.
\label{eqn:MSRDef}
\end{equation}

The degenerate eigenvectors at the EPD's are $\ket{R_{\pm i}} = \frac{1}{\sqrt{2}}\begin{pmatrix} \pm i \\ 1 \end{pmatrix}$. There are two types of exceptional point degeneracies labeled by $M_S^R = \pm i$, and they are constrained to live on curves of constant ${\cal R}e(M_S^R) = 0$. Note that the eigenvectors are now \textit{orthogonal} to each other only if ${\cal I}m(M_S^R) = 0$. Because there is now only one constraint on eigenvector orthogonality, the topology of the eigenvector orthogonality becomes curves (for two-dimensional parameter spaces) which partition the space in domains. In particular, the curves of eigenvector orthogonality delineate domains that separate the two types of EPD's. Here we identify a direct physical manifestation of this EPD charge through the CPA+EPD application discussed below. We also note that the scattering EPD's identified in this manner are shared by several closely-related matrix quantities, namely the impedance matrix $Z$, the Wigner reaction matrix $K$, the admittance $Y$, etc.

The complex functions $S_{\pm i}$ from Eqs. (\ref{Splus},\ref{Sminus}) take the simple form 
$S_{\pm i}^R \equiv S_{11} - S_{22} \mp 2iS_{21}$,
allowing one to conclude that there are four types of scattering EPD's in generic reciprocal scattering systems, labeled by $M_S^R = \pm i$ and by the winding number $n = \pm 1$. Every EPD is generically connected to exactly one other EPD by a finite curve defined by the conditions ${\cal R}e(M_S^R) = 0$ and $-1 \le {\cal I}m(M_S^R) \le 1$ (see discussion above).\\

\section{Experimental Results and Theoretical Modeling}
\label{Sec:EandT}
\begin{figure*}[htb]
\centerline{%
\includegraphics[width=18.4cm]{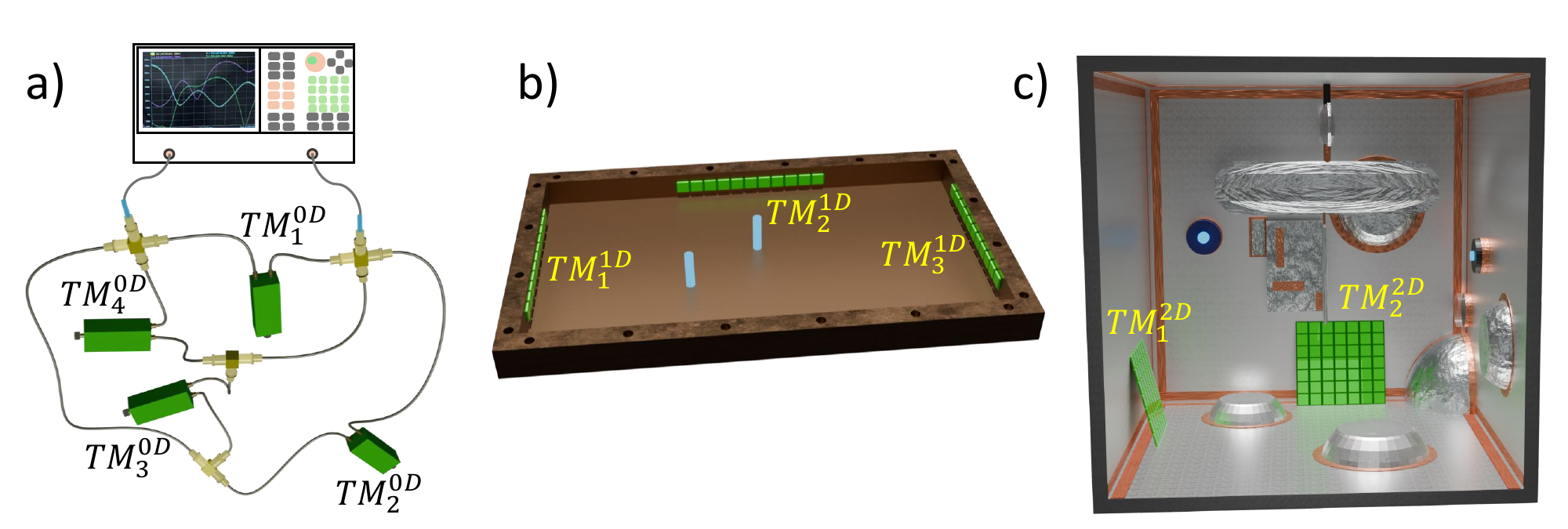}}
\caption{Schematics of three experimental systems of different wave-propagation dimension (1-3) used in this work. The tunable phase shifters and metasurfaces of each system are labeled by $TM_p^{qD}$, where $p$ labels each device within a system and $q$ indicates the dimension of the device. The ports connecting the vector network analyzer to each system are indicated with the light blue cylinders. a) Schematic view of a quasi-one-dimensional microwave graph. Along four of the bonds of the graph are tunable phase shifters ($TM_p^{0D}$) that serve to vary the electrical length of the bonds. b) Schematic view of a quasi-two-dimensional rectangular billiard with three tunable metasurfaces ($TM_p^{1D}$) along the walls of the billiard, with the top lid removed. c) Schematic view of a three-dimensional microwave cavity with two tunable metasurfaces ($TM_p^{2D}$) along the interior walls of the cavity, along with a mode-stirrer and irregular scatterers, with the front wall removed.}
\label{Schematics}
\end{figure*}

We start the analysis by characterizing the SU-EPD's of three classes of open resonant microwave cavities (see Appendix \ref{sec.Exp}): (a) quasi-one-dimensional quantum/microwave graphs/networks (Fig. \ref{Schematics}a); (b) quasi-two-dimensional microwave billiards (Fig. \ref{Schematics}b), and (c) three-dimensional microwave resonators with complex metallic boundaries (Fig. \ref{Schematics}c). All these systems have large dimensions compared to the electromagnetic wavelength, making them complex multi-modal scattering systems that are quite sensitive to small (sub-wavelength) perturbations. They contain multiple electronically-controlled phase-shifters (for graphs) and metasurfaces that modify the amplitude and phase of either propagating and/or reflecting waves. These are labeled as $TM_p^{qD}$, where $p$ labels each device within a system and $q$ indicates the dimension of the device. For phase shifters $q=0$, and for one and two-dimensional metasurfaces $q=1,2$ respectively. In the case of metasurfaces, the reflection phases and amplitudes of an impinging wave are modulated by a global voltage variation of the varactor diodes of the metasurface \cite{PhysRevApplied.20.014004,Erb24}. These tunable devices allow us to parametrically vary the geometric features of the isolated system, and therefore also the scattering matrix. In particular, the electronic control of the devices allows us to quickly and repeatably explore large regions of the parameter space without the need to mechanically modify the systems. Importantly, the metasurfaces and phase-shifters provide \textit{non-Hermitian} perturbations to the scattering systems, making them particularly effective at uncovering and manipulating scattering EPD's and other types of non-Hermitian spectral singularities. %For the sake of brevity, the presented experimental results will focus on the one- and two-dimensional systems.

In all cases below we attach two interrogating channels in the scattering setups, resulting in a $2\times 2$ scattering matrix. The scattering parameters have been measured for various frequencies $f$ and perturbation parameters using a Keysight model N5242B microwave vector network analyzer (VNA) which is connected to the two ports of the scattering system through coaxial cables that support a single propagating mode. The VNA and cables leading to the system ports are calibrated by means of a Keysight model N4691D ECal Module, and the complex scattering matrix $S$ is measured over a chosen frequency and parameter range.

\subsection{Reciprocal scattering processes}

\begin{figure*}[htb]
\centerline{%
\includegraphics[width=18cm]{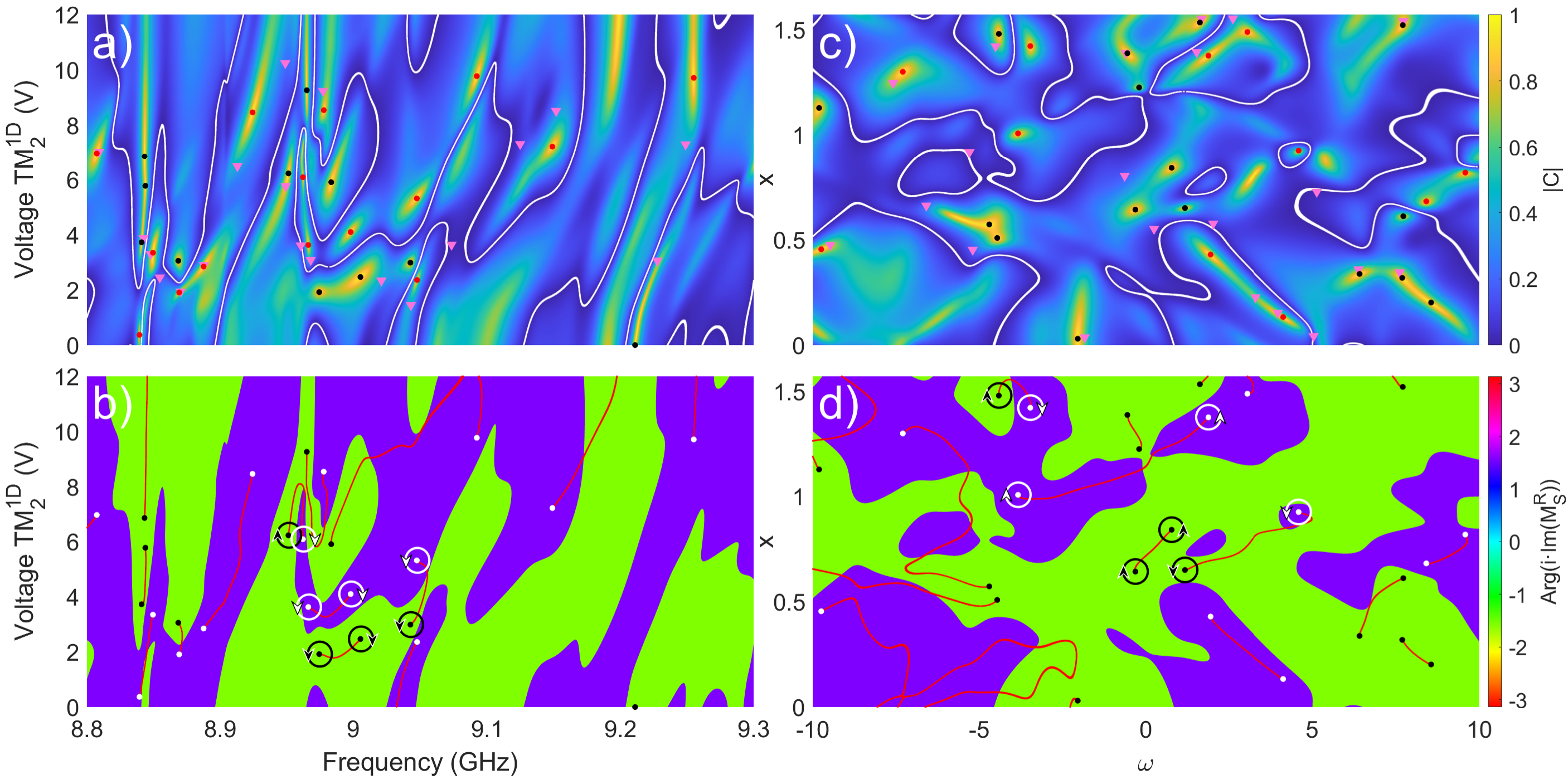}}
\caption{Exceptional point degeneracies, orthogonality curves, EPD domains, and EPD pair connections in a two-dimensional parameter space. a) Experimental eigenvector coalescence $|C|$ vs frequency and $TM_2^{1D}$ applied bias voltage in the rectangular billiard. b) $Arg(i\cdot {\cal I}m(M_S^R))$ vs frequency and $TM_2^{1D}$ applied bias voltage for the same data as in (a). c) Eigenvector coalescence $|C|$ vs $x$ and $\omega$ from the RMT cavity model. d) $Arg(i\cdot {\cal I}m(M_S^R))$ vs $x$ and $\omega$ for the simulation shown in (c). The red/white and black dots correspond to the $M_S^R=+i$ and $M_S^R=-i$ EPD's, respectively. The extended white regions are points of near zero eigenvector coalescence ($|C|<0.002$). The purple triangles correspond to points where $det(S)=0+i0$, which enable coherent perfect absorption. The different domains hosting the two types of EPD's ($\pm i$) are clearly shown in this plot. The green regions have a phase of $-\pi/2$ ($-i$ EPD's can live here) and the purple regions have a phase of $+\pi/2$ ($+i$ EPD's can live here). The zeros of ${\cal I}m(M_S^R)$ form the interface between the two different domains, and also correspond to the curves of orthogonality in the eigenvector coalescence $|C|$. The curves of ${\cal R}e(M_S^R) = 0$ and $-1 \le {\cal I}m(M_S^R) \le 1$ connecting EPD's are highlighted in red. Each EPD type can have a $\pm 1$ winding number, and EPD's connected by ${\cal R}e(M_S^R) = 0$ and $-1 \le {\cal I}m(M_S^R) \le 1$ have a particular relative winding number depending on their relative charge. The black and white circles and arrows indicate the winding number of the $-i$, $+i$ EPD's respectively, for a subset of the EPD's. }
\label{Jul12_Coalescence}
\end{figure*}

First, we analyze the formation of the scattering EPD's in the two-parameter space of frequencies and global voltage variation of $TM_2^{1D}$ for the two-dimensional cavity of Fig. \ref{Schematics}b. In Fig. \ref{Jul12_Coalescence}a we consider the reciprocal scenario and report, as a heat map, the measured coalescence $|C|$ which has been evaluated after diagonalizing the experimental scattering matrix. The same parametric analysis has been performed for the three-dimensional cavity of Fig. \ref{Schematics}c and for the network system of Fig. \ref{Schematics}a where the interference variations are performed by phase shifters placed along the coaxial cables (bonds) of the network. We see that the parameter space is partitioned in domains which are separated by orthogonality curves (see white curves) characterized by $|C|=0$. Surprisingly, there is a complete orthogonal set of eigenvectors along this continuous locus of points in parameter space, despite the non-Hermitian nature of this system. In each of the domains delineated by the $|C|=0$ curves, we have identified a number of EPD's corresponding to $|C|=1$ (red and black filled circles). In practice, the orthogonality and EPD conditions incorporate a tolerance for purposes of visualization.

A further distinction of positive-charged $M_S^R=+i$ (red-filled circles) and negative-charged $M_S^R=-i$ (black-filled circles) EPD's is done by analyzing the complex zeros of the functions $S_{\pm i}^R$ (Eqs. (\ref{Splus},\ref{Sminus}) evaluated for the special case of reciprocal transport). In Fig. \ref{Jul12_Coalescence}b we report the same data of Fig. \ref{Jul12_Coalescence}a by referring to the variable $Arg(i\cdot {\cal I}m(M_S^R))$. The latter takes values $\pm \pi/2$; thus highlighting in a better manner the partition of the parameter space in positive-charged ($+\pi/2$; purple areas) and negative-charged ($-\pi/2$; green areas) domains. The boundaries between the domains are the curves of eigenvector orthogonality (${\cal I}m(M_S^R) = 0$). In this figure, we indicate for some example cases the winding number evaluated after the calculation of the path integral in Eq. (\ref{eqn:winding_num}). Furthermore, we have plotted with red the finite curves that connect pairs of EPD's with opposite charges (winding number) and the same winding number (charge). These curves, as discussed above, are characterized by the constraints ${\cal R}e(M_S^R)=0$ and $-1 \le {\cal I}m(M_S^R) \le 1$. 

The rectangular billiard of Fig. \ref{Schematics}b is a pseudointegrable billiard \cite{Zyczkowski91}, but all the results discussed above are also present for the fully ray-chaotic quarter bow-tie billiard \cite{Erb24} (see Appendix \ref{sec.Reciprocal_Systems}). Let us point out that the same picture emerges also in the case of reciprocal quasi-one dimensional networks (see Fig. \ref{Schematics}a) and complex three-dimensional cavities with underlying chaotic ray dynamics (see Fig. \ref{Schematics}c), shown in Appendix \ref{sec.Reciprocal_Systems}.

The applicability of the topological rules that characterize the EPD's can be further confirmed theoretically, using Random Matrix Theory (RMT) modeling. The latter is established as an appropriate model for the statistical description of the universal scattering properties of complex cavities such as reverberant chambers; relevant to realistic wireless communication applications \cite{Stock99,Haake10}. To this end, we have devised an appropriate RMT model (see Appendix \ref{sec.RMT}) whose scattering matrix takes the form

\begin{equation}
\begin{aligned}
\label{RMTS}
S(\omega) &= -1_2 + i W G(\omega)W^T,\\ 
G(\omega) &= \frac{1}{\omega - (H_0 - \frac{i}{2} W^T W)},
\end{aligned}
\end{equation}
where $W$ is the $2\times N$ coupling matrix
that connects the two interrogating channels with two resonant modes of an $N$-dimensional effective Hamiltonian $H_0$ that models a cavity with additional degrees of freedom, such as metasurfaces.
The Hamiltonian $H_0$ that describes the isolated cavity (with the metasurfaces) takes the form:
\begin{equation}
H_0(x,y) = H_1 + \lvert cos(x) \rvert \lvert cos(y) \rvert H_2 + \lvert sin(x) \rvert \lvert sin(y) \rvert H_3.
\label{eqn:H0Def}
\end{equation}
The Hamiltonian $H_1$ belongs to the Gaussian Orthogonal Ensemble (GOE) with matrix elements given by a Gaussian distribution of mean zero and standard deviation $\sqrt{\frac{N}{\pi}}$. Similarly, the off-diagonal elements of $H_2$ and $H_3$ are taken from a GOE of the same family as $H_1$, while the diagonal elements are taken from the uniform distribution $[0,-i]$. The trigonometric functions present in Eq. (\ref{eqn:H0Def}) serve to perturb the system such that the norm of the Hamiltonian is conserved. The added degrees of freedom of $H_0$ make the Green's function and scattering matrix functions of $x$, $y$, and $\omega$, with $x$ and $y$ acting as surrogates for the tunable metasurfaces in the networks and cavities, allowing us to better understand the scattering properties of such systems.

In Figs. \ref{Jul12_Coalescence}c,d we report the heat map for the variables $|C|$ and $Arg(i\cdot {\cal I}m(M_S^R))$, respectively for the RMT model. The emergence of a similar picture to the one found in the experimental analysis is evident. Since the RMT model has no geometrical or hidden symmetries, and represents a fully generic resonant system, this re-confirms the validity of our theoretical description of scattering EPD's.

\subsection{Non-reciprocal scattering processes} 

\begin{figure}[htb]
\centerline{%
\includegraphics[width=8.7cm]{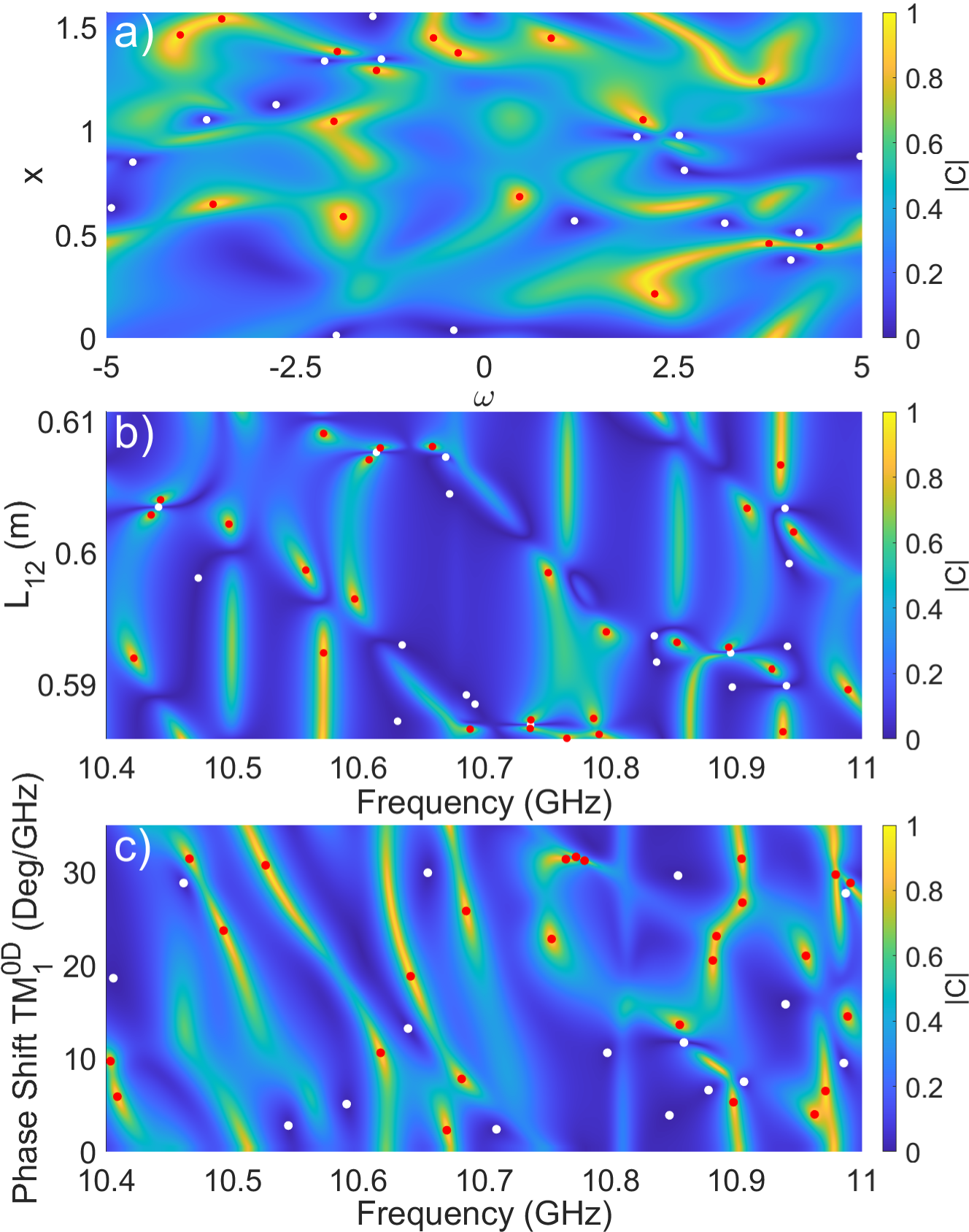}}
\caption{Exceptional point degeneracies and orthogonality points in two-dimensional parameter spaces in non-reciprocal scattering systems. The white dots correspond to orthogonality points (OP's) and the red dots correspond to EPD's. a) Eigenvector coalescence $|C|$ vs $\omega$ and $x$ for the RMT cavity model. b) Eigenvector coalescence $|C|$ vs frequency and bond length $L_{12}$ for the analytic model of a tetrahedral graph. c) Eigenvector coalescence $|C|$ vs frequency and phase shift of $TM_1^{0D}$ of the experimental tetrahedral graph.}
\label{Non_Recip_Graph}
\end{figure}

We have argued above that a constraint for the existence of orthogonality curves is the presence of reciprocity. When the latter is violated, we expect the disintegration of the orthogonality curves, which must turn to discrete points in the two-dimensional parameter space. To validate this prediction we have modified the RMT model of Eqs. (\ref{RMTS},\ref{eqn:H0Def}) in a way that incorporates the effects of a magnetic field \cite{Beenakker97}. Specifically, we have substituted the Hamiltonian $H_0$ of the isolated cavity by 
\begin{equation}
H_{mag} = H_0 + i \alpha B,
\label{eqn:H_NRDef}
\end{equation}
where $B=-B^T$ is an anti-symmetric matrix with matrix elements given by a normal Gaussian distribution. The parameter $\alpha$ models the relative strength of the magnetic field. In Fig. \ref{Non_Recip_Graph}a we report the heat map for the coalescence $|C|$ for a non-zero magnetic field corresponding to $\alpha=1$ (fully broken time-reversal symmetry). Indeed, we find that the orthogonality boundaries that separate the different EPD-charges have been destroyed and replaced with discrete orthogonality points. Nevertheless, the topological features of $M_S=\pm i$ and $n=\pm 1$ of each EPD are maintained separately.

We have further validated the destructive effect of non-reciprocal behavior by analyzing the parametric response of the EPD's for the complex network of coaxial cables, see Fig. \ref{Schematics}a, which emanate from $N$ microwave T-junctions (vertices). The cables connecting the vertices $\nu$ and $\mu$ have irrational lengths $L_{\nu\mu}$. The two varied parameters of this physical system are the frequency $f=(c_0/2\pi n^{(r)})k$ ($k$ is the wavenumber, $c_0$ is the speed of light and $n^{(r)}$ is the complex-valued relative index of refraction that models the losses of the coaxial cables $L_{\nu\mu}$) of the injected wave and the length of one bond. The theoretical analysis of this system follows along the lines indicated in Ref. \cite{Kottos00,Kottos03,Wang24}. The corresponding scattering matrix takes a form similar to the one describing the RMT model (see Appendix \ref{sec.Networks} for the theoretical modeling of the scattering matrix). Specifically,
\begin{equation}
S = -{\hat 1} + 2i W \frac{1}{h + i W^T W} W^T,
\end{equation}
where $W$ is the $2\times N$ coupling matrix that describes the coupling between the interrogating channels and two vertices of the networks. The matrix elements are $W_{\beta,\nu}=\delta_{\beta,\nu}$ where $\beta=1,2$ is the channel-index and $\nu=1,\dots,N$ is the vertex index of the network. The matrix $h$ incorporates information about the connectivity and metric characteristics of the graph and takes the form

\begin{equation}
h_{\nu \mu}(k) = 
\begin{cases}
-\sum_{l\neq \nu} {\cal A}_{\nu l}cot(k L_{\nu l}), & \nu=\mu \\
{\cal A}_{\nu \mu}csc(k L_{\nu \mu})\cdot e^{i \phi_{\nu \mu}}, & \nu\neq \mu
\end{cases}
\label{eqn:MDef}
\end{equation}
where ${\cal A}$ is the adjacency matrix having elements ${\cal A}_{\nu\mu}=1$ if the vertices $\nu,\mu$ are connected via a bond and zero otherwise and $\phi_{\nu\mu}= -\phi_{\mu\nu}$ is an effective magnetic flux. For simplicity, we assume that the magnitude of $\phi_{\nu\mu}=\phi_0$ is uniform throughout all bonds of the graph. The modeling of Lorentz reciprocity symmetry violation via the magnetic flux in Eq. (\ref{eqn:MDef}) doesn't exactly reproduce the effects of microwave circulators that have been used in our experiments. Nevertheless, the effect of reciprocity violation, i.e. the destruction of topological domains, is generically present (see also the results of the RMT modeling Fig. \ref{Non_Recip_Graph}a). For one of the simplest analytic models that exhibits EPD's and their dynamics, see Appendix \ref{sec.Simple_Network}.

In Fig. \ref{Non_Recip_Graph}b we report the coalescence $|C|$ vs frequency $f$ and bond length $L_{12}$ for an analytic tetrahedral graph in the presence of a magnetic field $\phi_0\neq 0$. In the scanned parameter domain we find numerous EPD's. However, the orthogonality curves have now turned into discrete points in complete analogy with the RMT results. We remind the reader that the eigenvector orthogonality requires that the two conditions ${\cal I}m(M_{S}) = 0$ and $\left| \frac{S_{12}}{S_{21}} \right| = 1$ have to be satisfied simultaneously. These two constraints define two sets of curves (in a two-dimensional parameter space). They are simultaneously satisfied at discrete points which correspond to their intersections. 

The corresponding experimental results for $|C|$ are shown in Fig. \ref{Non_Recip_Graph}c for a tetrahedral microwave graph. Experimentally, the variations of the bond-lengths are controlled using a tunable phase shifter ($TM_p^{0D}$). In this platform, reciprocity is violated by replacing a reciprocal SMA T-junction with a 3-port non-reciprocal microwave circulator. This data confirms again that when reciprocity is violated the orthogonality curves turn into points and thus, the topological domains of EPD charges are destroyed.

\subsection{Parametric dynamics and creation/annihilation of EPD's and OP's} 

Let us finally comment on the possibility to manipulate (create/annihilate) the scattering singularities. The process requires the expansion of the dimensionality of the parametric space by employing variations of a third parameter of the scattering system. For our systems, the third variational parameter is chosen to be another tunable phase shifter (1D) or the applied bias voltage of another metasurface (2D/3D). 

Based on the theoretical analysis we expect that the creation and annihilation of the EPD's via their collisions will involve singularities of the same charge. Therefore the dynamics will be bounded by the orthogonality curves i.e. EPD's belonging to different domains (separated by an orthogonality boundary) will not interact with one another. Furthermore, these collisions and creation processes will conserve the total winding number of the pair involved in the process. In other words, only a pair of EPD's of the same charge and opposite winding number can be annihilated (via parametrically-induced collisions) or created.

This theoretical expectation is confirmed nicely from experimental data of the two-dimensional cavity (Fig. \ref{Schematics}b), see Fig. \ref{Aug16_Movie} (for a video of variation of the third control parameter see Appendix \ref{sec.EPD_Creation}). In this figure we report two specific scenarios: Subfigures \ref{Aug16_Movie}(a-b) show the creation of a pair of EPD's with opposite winding numbers occurring in the positive charge ($M_S^R=+i$) domain as we vary the third parameter. Similarly, subfigures \ref{Aug16_Movie}(c-d) indicate the collision (and consequent annihilation) of two positive charged EPD's with opposite winding numbers. This event occurs via the motion of the two EPD's in the positive charged domain. The creation and annihilation rules of EPD's as detailed above have also been confirmed via the RMT model of the previous section.

\begin{figure}[htb]
\centerline{%
\includegraphics[width=8.7cm]{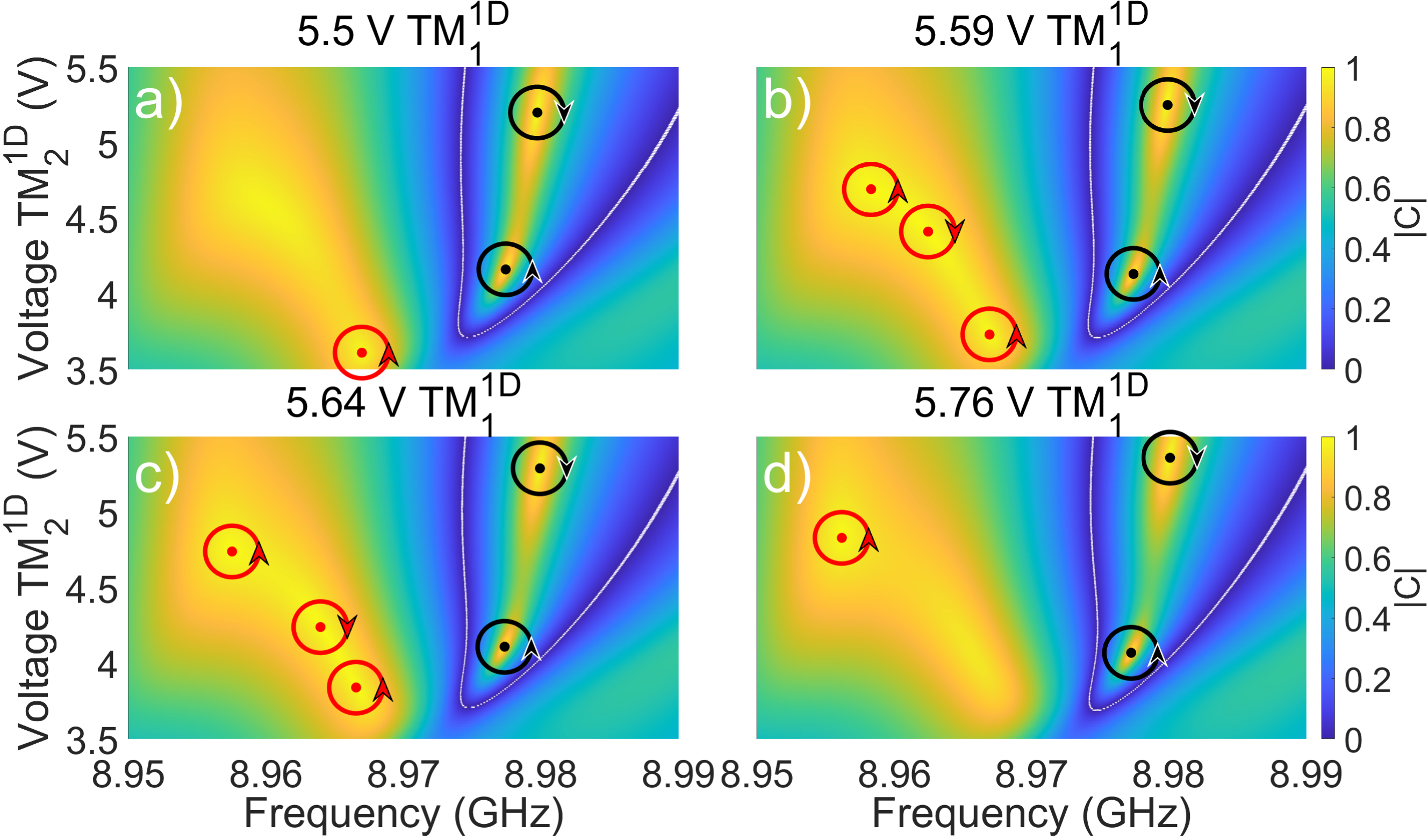}}
\caption{Illustration of exceptional point degeneracy creation and annihilation events. Eigenvector coalescence $|C|$ vs frequency and $TM_2^{1D}$ applied bias voltage in the rectangular billiard. Each of the four plots (a-d) are at a different fixed applied bias voltage of $TM_1^{1D}$. a)-b) Shows a creation event of two $M_S^R=+i$ EPD's of opposite winding number, while c)-d) shows the annihilation of two $M_S^R=+i$ EPD's of opposite winding number. The white regions are points of near zero eigenvector coalescence ($|C|<0.005$) and the red and black dots correspond to the $M_S^R=+i$ and $M_S^R=-i$ EPD's, respectively. The red and black circular arrows indicate the winding numbers of the exceptional point degeneracies.}
\label{Aug16_Movie}
\end{figure}

In the following videos of this section, the parametric dynamics of the various singularities discussed in this work are directly illustrated. As shown above, the results for reciprocal and non-reciprocal systems are generic and independent of specific system details. In Video \ref{vid.Reciprocal}, we show the dynamics of $|C|$ in the rectangular billiard (Fig. \ref{Schematics}b) as we vary the third parameter of the system. The EPD's are marked with red ($M_S^R=+i$) and black ($M_S^R=-i$) dots, and the orthogonality curves are marked in white. In this video, there are numerous creation and annihilation events of EPD pairs of the same charge, but otherwise each individual EPD is topologically stable in the parameter space. Each EPD charge stays within their EPD domain formed by the curves of orthogonality and can never interact with an EPD of the opposite charge. However, EPD's within one domain can travel into a different domain of the same charge, as seen by the orthogonality curves which can connect or disconnect different domains of the same EPD charge.

\begin{video}
\hspace*{-0.5cm}
\includegraphics[width=9.05cm]{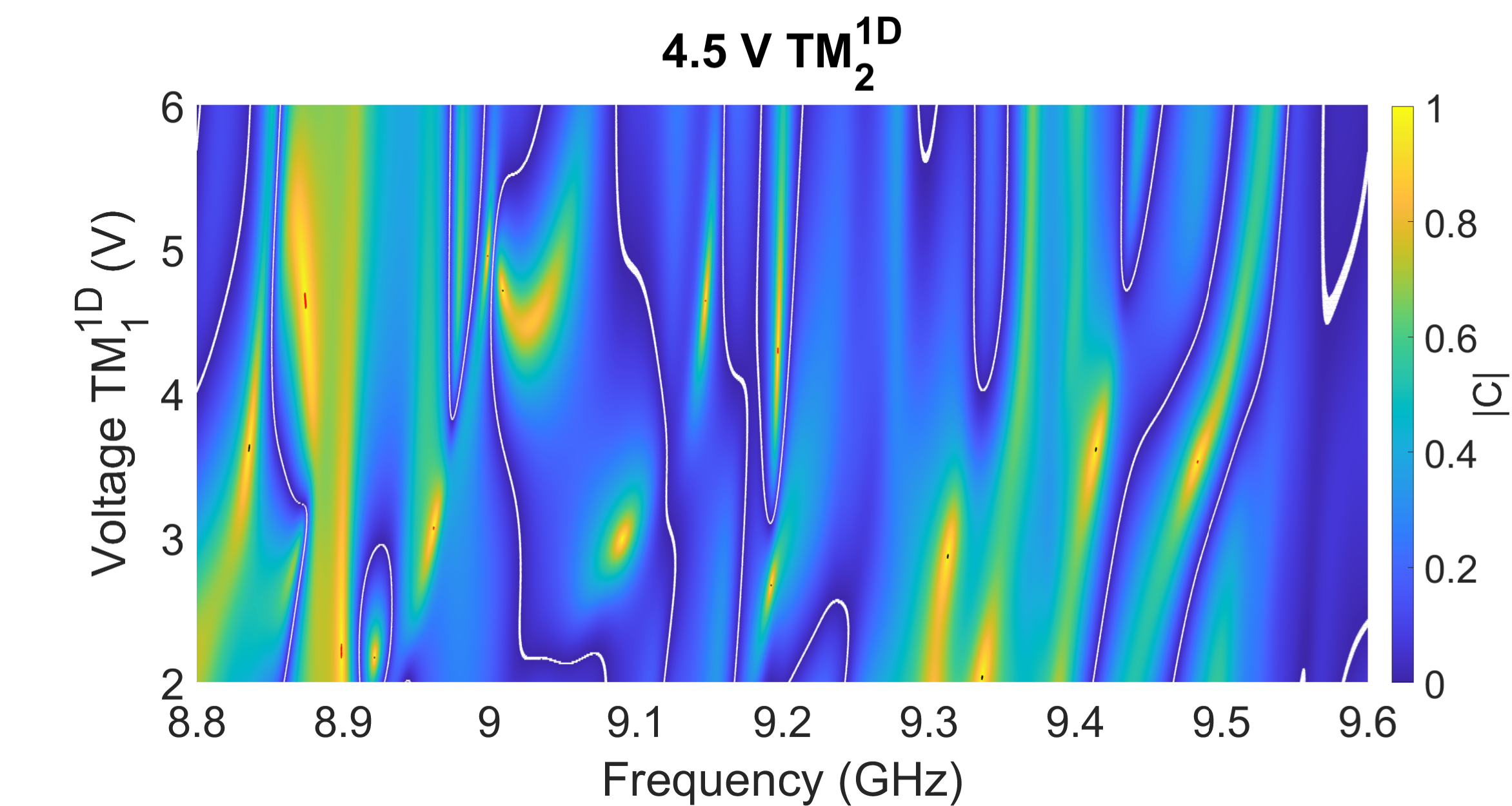}
%\setfloatlink{http://some.video.com/fun.mov}
\caption{\label{vid.Reciprocal} Dynamics and interactions of exceptional point degeneracies and orthogonality curves as a third parameter of the system is varied. The image above shows the eigenvector coalescence $|C|$ vs frequency and $TM_1^{1D}$ applied bias voltage in the experimental rectangular billiard for the first frame of the video. Each frame of the video is at a different fixed applied bias voltage of $TM_2^{1D}$. In the video the $\pm i$ EPD's are highlighted by the red and black points where $|{\cal R}e(M_S^R)| \le 0.0085$ and $|{\cal I}m(M_S^R) \mp 1| \le 0.0085$, and the white eigenvector orthogonality curves are marked by $|C| \le 0.003$. In the video we see numerous EPD creation and annihilation events as the applied bias voltage of $TM_2^{1D}$ varies.}
\end{video}

In Video \ref{vid.Domains}, the same underlying data as in Video \ref{vid.Reciprocal} is shown, but now plotted in a way to highlight the EPD domains and the EPD connection curves. The EPD's are marked with white ($M_S^R=+i$) and black ($M_S^R=-i$) dots, the EPD domains are the purple and green regions, and the EPD connection curves are marked by the red curves. In this video, the dynamics of the boundaries of the EPD domains, and how different domain regions can combine or separate, is particularly clear. The EPD connection curves are also manipulated as the third parameter is varied and the connections between two EPD pairs can exchange partners, but all EPD pair connection curves still follow the winding number-charge relation specified in Section \ref{Sec:SandEPD}. A closed loop of orthogonality (i.e. an EPD domain) can also shrink down into nothing and disappear, or be created and grow inside an opposite-charge domain. For a more detailed view of a portion of the video see Appendix \ref{sec.Domains}. 

\begin{video}
\hspace*{-0.55cm}
\includegraphics[width=9.1cm]{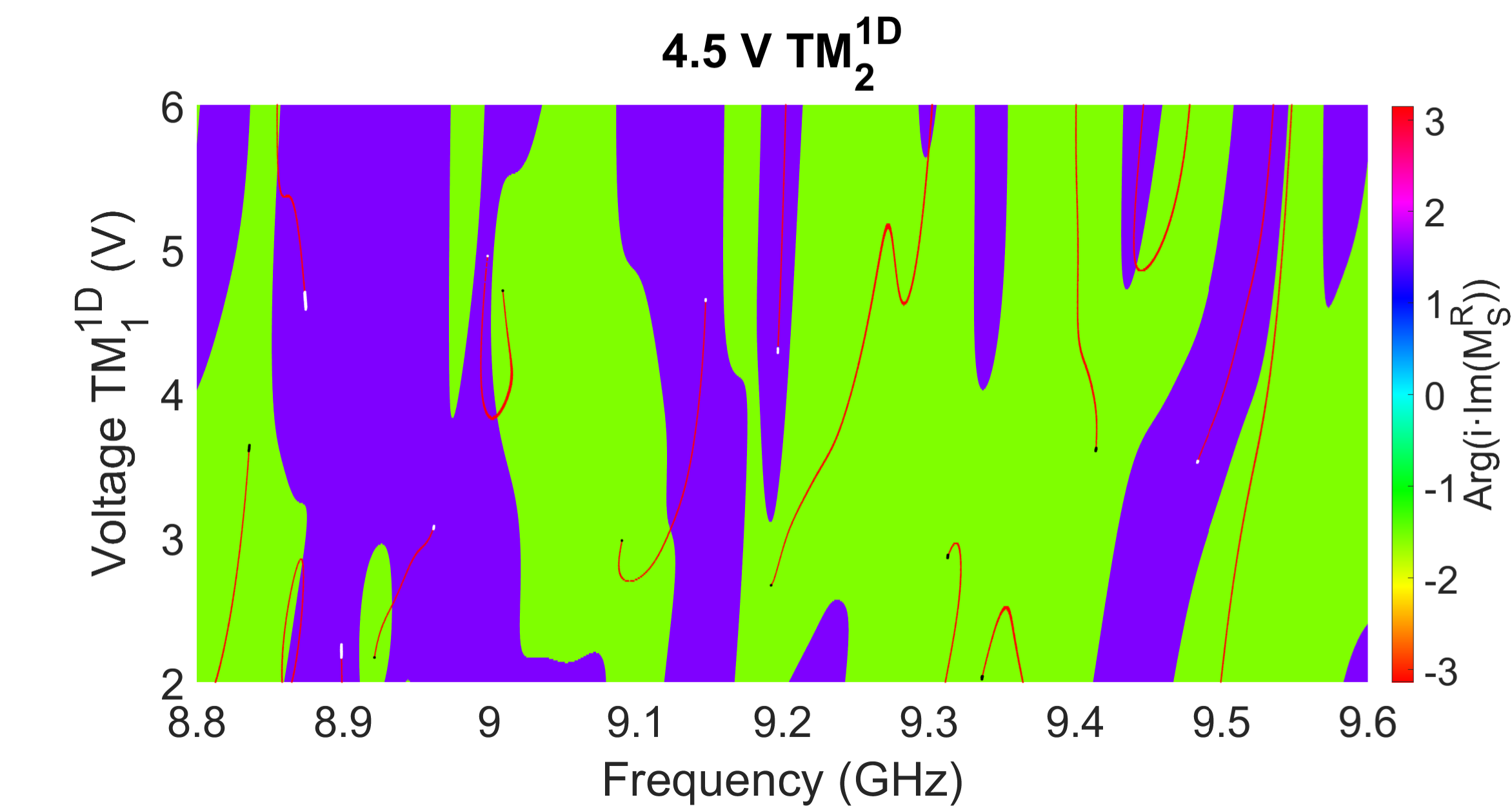}
%\setfloatlink{http://some.video.com/fun.mov}
\caption{\label{vid.Domains} Dynamics and interactions of EPD domains and EPD connection curves as a third parameter of the system is varied. The data shown in this figure is exactly the same as shown in Video \ref{vid.Reciprocal}, but plotted in a different way. The image above shows $Arg(i\cdot {\cal I}m(M_S^R))$ vs frequency and $TM_1^{1D}$ applied bias voltage in the experimental rectangular billiard for the first frame of the video. Each frame of the video is at a different fixed applied bias voltage of $TM_2^{1D}$. In the video the $\pm i$ EPD's are highlighted by the white and black points where $|{\cal R}e(M_S^R)| \le 0.01$ and $|{\cal I}m(M_S^R) \mp 1| \le 0.01$, and the red EPD connection curves are marked by $|{\cal R}e(M_S^R)| \le 0.003$ and $-1 \le {\cal I}m(M_S^R) \le 1$. The domains that the two types of EPD's are allowed to live in are indicated by the green ($-i$) and purple ($+i$) regions. In the video we clearly see the dynamics of the EPD domains and the reconnection events of the EPD connection curves between different EPD pairs.}
\end{video}

In Video \ref{vid.Non_Reciprocal}, the dynamics of $|C|$ for the non-reciprocal experimental tetrahedral graph (Fig. \ref{Schematics}a) are shown. The EPD's are marked with red dots and the orthogonality points are marked with white dots. Distinctly different from Video \ref{vid.Reciprocal}, the curves of orthogonality are now points in this non-reciprocal system. These points show dynamics that are qualitatively similar to that of the EPD's, including continuous evolution in parameter space, and pair-creation and annihilation events conserving winding number. In this video, we see an abundance of creation and annihilation events of both the EPD's and orthogonality points. As in the reciprocal case, the curves connecting pairs of EPD's can connect or disconnect different pairs of EPD's (see Appendix \ref{sec.NR_EPD_Connections}).

\begin{video}
\hspace*{-0.44cm}
\includegraphics[width=8.9cm]{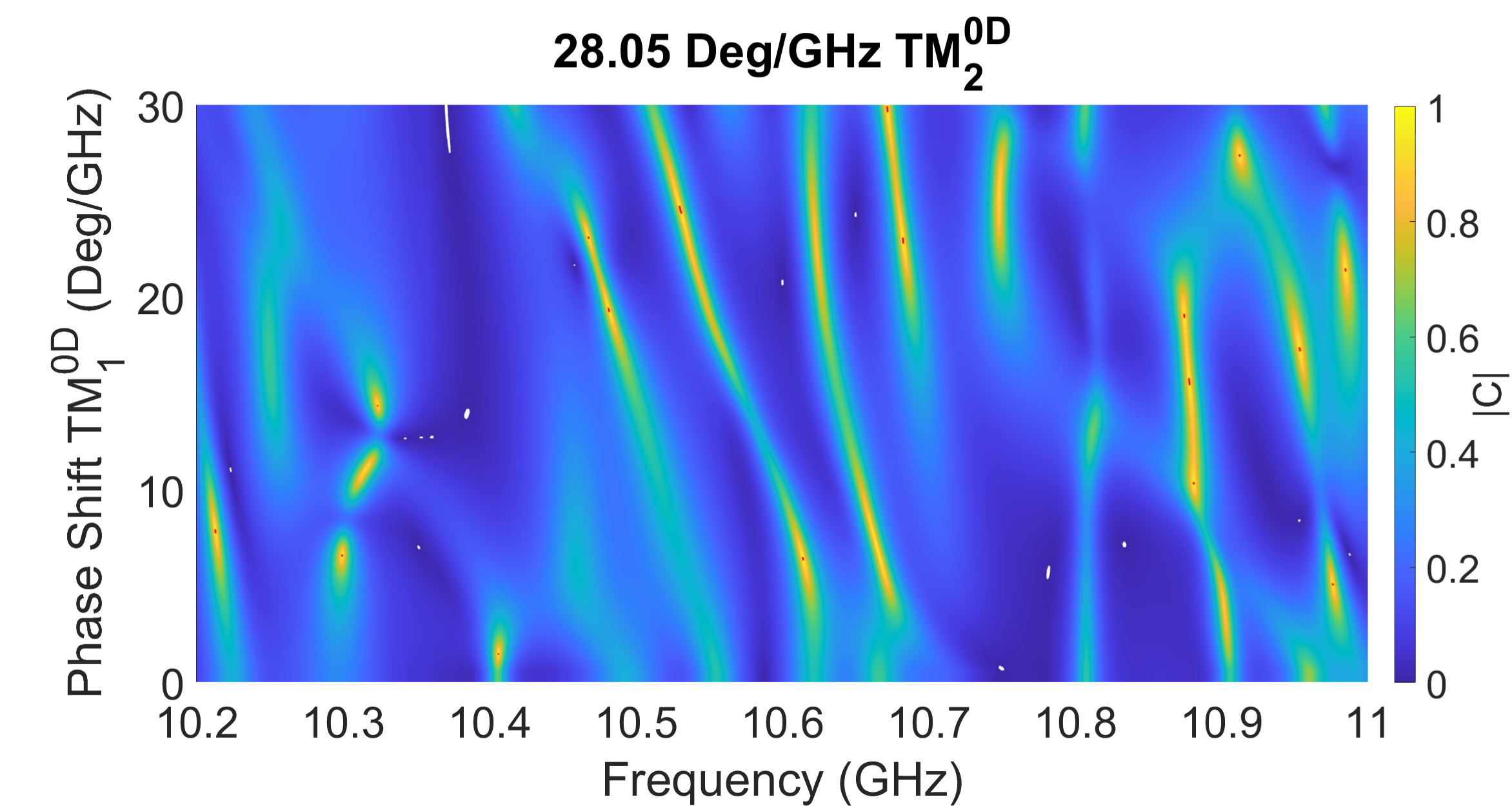}
%\setfloatlink{http://some.video.com/fun.mov}
\caption{\label{vid.Non_Reciprocal} Dynamics and interactions of exceptional point degeneracies and orthogonality points as a third parameter of the system is varied. The image above shows the eigenvector coalescence $|C|$ vs frequency and $TM_1^{0D}$ phase shift for the experimental tetrahedral graph with non-reciprocity for the first frame of the video. Each frame of the video is at a different fixed phase shift of $TM_2^{0D}$. In the video the EPD's are highlighted by the red points where $|C| \ge 0.985$, and the white eigenvector orthogonality points are marked by $|C| \le 0.002$. In the video we see numerous EPD and OP creation and annihilation events as the phase shift of $TM_2^{0D}$ varies.}
\end{video}

\section{Application: Robust 50:50 I/Q splitter}
\label{Sec:Robust5050}
The possibility to parametrically control the formation of EPD's by manipulating the geometric and dynamical (e.g. reciprocity) symmetries of a cavity offers new opportunities for the control of wave scattering processes. An example case is the realization of a robust 50:50 power splitter that outputs equal intensity waves between the two ports of the system regardless of the magnitudes and phases of the input waves. In fact, the outgoing signal from the proposed splitter maintains a fixed $\pi/2$ relative phase between the two monochromatic waves that propagate along the two interrogating channels, making it a valuable source of in-phase/quadrature (I/Q) signals.

The starting point for the implementation of such a splitter is the observation that under the reciprocal condition $S_{12}=S_{21}$, the degenerate eigenvectors at the scattering EPD conditions take the form $\ket{R_{\pm i}} = \frac{1}{\sqrt{2}}\begin{pmatrix} \pm i \\ 1 \end{pmatrix}$ (see Eq. (\ref{DV}) and below Eq. (\ref{eqn:MSRDef})). The corresponding degenerate eigenvalue can be evaluated from Eq. (\ref{eval}) by substituting $S_{12}=S_{21}$, while the Jordan vector $\ket{J_{\pm i}}$ is given in Equation \ref{Jvec_rec}.

We consider a generic monochromatic wave $\ket{\alpha_{\rm in}}$ being injected in the scattering domain which has been tuned to be at EPD conditions. In this case, the incoming wave can be expressed as $\ket{\alpha_{\rm in}} = c_1 \ket{R_{\pm i}} + c_2 \ket{J_{\pm i}}$. The corresponding outgoing signal is then $\ket{\alpha_{\rm out}}=S \ket{\alpha_{\rm in}} = \lambda_S^{\pm i} \left[c_1 \ket{R_{\pm i}} + c_2 \ket{J_{\pm i}}\right] + c_2 \ket{R_{\pm i}}$. From this last expression, it is clear that if the degenerate eigenvalue $\lambda_S^{\pm i}$ happens to be zero, the outgoing wave will take the simple form $\ket{\alpha_{\rm out}}= c_2 \ket{R_{\pm i}}$. In other words, under the combined requirements of EPD condition and null scattering eigenvalue we come up with an output signal that has equally split channel amplitudes and a relative phase of $\pi/2$ between the two outgoing propagating waves at each of the two channels. The combined conditions for such a case are:
\begin{equation}
\begin{cases}
S_{11} = -S_{22}\\
S_{11} = \pm i S_{21}.
\end{cases}
\end{equation}

The requirement of degenerate zero eigenvalue corresponds to the Coherent Perfect Absorption (CPA) exceptional point singularity \cite{Stone19,Yang21,Farhi22}. Coherent Perfect Absorption is the scenario where a \textit{specific} injected wavefront -- associated with the corresponding eigenvector of the scattering matrix -- is trapped via destructive interferences inside the cavity and completely absorbed. The paradoxical situation is resolved by realizing that this is a zero-measure injected wavefront, while the 50:50 output split occurs for any other generic incident wave.

Naively, one would expect that a system with degenerate zero scattering eigenvalues would absorb all input energy. However, this is not the case because the Jordan vector $\ket{J_{\pm i}}$ provides a route by which energy can emerge from the system despite being at a CPA+EPD condition. The total magnitude of the outgoing signal directly depends on the weight $c_2$ of $\ket{\alpha_{\rm in}}$, allowing us to control the total output power of the 50:50 beam splitter.

\begin{figure*}[htb]
\centerline{%
\includegraphics[width=17.9cm]{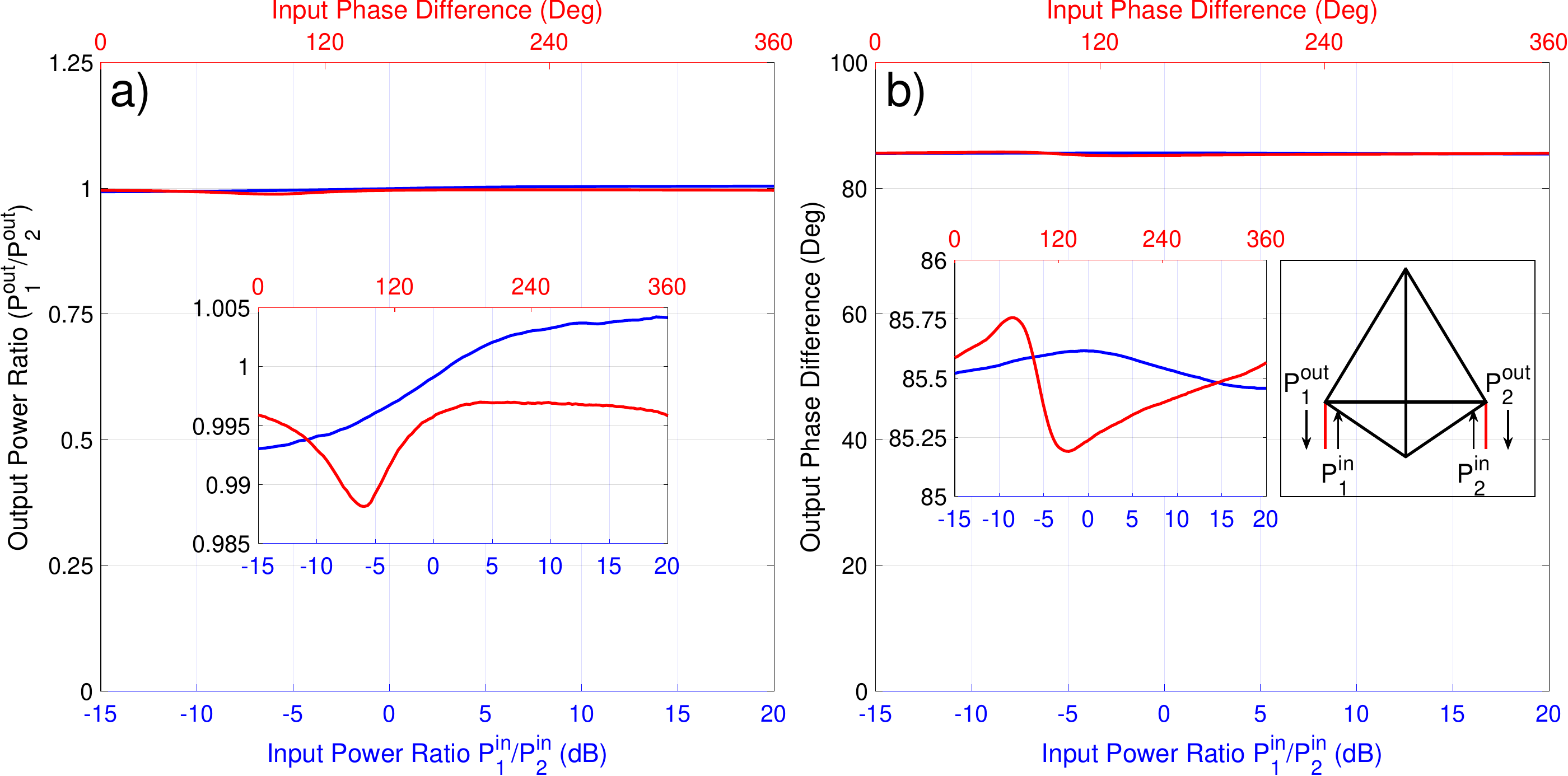}}
\caption{Experimental demonstration of the robust 50:50 I/Q splitting of CPA+EPD in a reciprocal tetrahedral microwave network. The frequency for the measurement of the $+i$ EPD+CPA was fixed at approximately 9.36 GHz. a) Output power ratio vs input power ratio (lower axis, blue) and input phase difference (upper axis, red). The blue curves correspond to the input signals power ratio being swept, and the red curves correspond to the input signals relative phase being swept. Inset shows a vertical zoom-in to illustrate the details of the curves. b) Output phase difference vs input power ratio (lower axis, blue) and input phase difference (upper axis, red). The blue and red curves have the same interpretation as in (a). Left inset shows a vertical zoom-in to magnify the details of the curves. Right inset depicts the input and output signals to/from the graph.}
\label{CPA_EP_Recip}
\end{figure*}

We have experimentally tested the above proposal using a reciprocal one-dimensional tetrahedral microwave network. The parametric variations of four lengths of the network have been performed using four phase shifters. The additional varying parameter was the frequency of the injected wave. All these parameters were changed until a condition was found where the system nearly satisfied the CPA and EPD conditions simultaneously. The measured outgoing signals at each channel have been collected and analyzed for a range of injected waves with different channel amplitudes and relative phases. The results shown in Fig. \ref{CPA_EP_Recip} demonstrate a robust 50:50 outgoing signal with (rigid) $\approx\pi/2$ relative phases. The results show small deviations from the ideal amplitude ratio and phase difference because the scattering eigenvalues showed $|\lambda_S^{1}|+|\lambda_S^{2}|=0.054 \neq 0$.  For further description of the measurement and injection procedure, and some representative examples of arbitrary signals (not at a CPA or EPD condition) sent into the system, see Appendix \ref{sec.Verification}.

\section{Discussion}
\label{Sec:Diss}
With the ability to find and manipulate numerous scattering singularities in generic systems by utilizing tunable parameters, a much more thorough exploration of their exotic behavior and topological properties is possible. Having tunable parameters embedded in the system removes the need to do any special design or engineering of a system to have particular scattering singularities at certain frequencies, which greatly increases their accessibility and ubiquity. Their topological protection in a two-dimensional parameter space (as long as they are not annihilated with a singularity of the opposite winding number) makes applications of these singularities much more feasible.

We have demonstrated that pairs of singularities can be created or annihilated, conserving winding number. We are also able to explore the topological properties of the scattering singularities and understand their interactions based on the zeros of complex scalar functions \cite{Nye74,Neu90,Shvartsman94,Berry00}. (We note in passing that prior work has examined topological defects in superpositions of vector waves in real space, revealing a number of complex features \cite{Hajnal87,Hajnal90,Berry03,Berry09}) By converting components of the scattering matrix into various complex scalar functions, we are able to describe many of the singularities and features of the experimental data as the zeros of these functions. This approach reveals the underlying mathematical structure of scattering singularities and establishes general organizing principles for understanding the scattering processes occurring in complex systems. 

We introduce a random matrix theory-based and analytic model with tunable parameters that exhibit the same scattering singularities and their dynamics and interactions that are present in our experimental data. From these models, it is clear that an abundance of scattering singularities can be found in any complex wave propagation system with tunable parameters. We expect that these models will motivate a greater focus on the study of the sub/super-unitary scattering matrix in addition to non-Hermitian Hamiltonians, as there are a variety of rich physical phenomena and topological properties that can be more easily explored using the scattering matrix.

The $M_S$, $M_S^R$ complex scalar functions we introduce, easily constructed from raw data, contain a wealth of information about the scattering singularities and features seen in the experimental data. We are able to show that there are two distinct charges of exceptional point degeneracies, each with an associated winding number, and that they are confined to distinct domains of parameter space for reciprocal systems. We identify for the first time the curves of eigenvector orthogonality in reciprocal systems, the orthogonality points in non-reciprocal systems, and restricted domains where the exceptional point degeneracies are permitted to exist in parameter space. We also show how pairs of EPD's are connected with one another and how this connection dictates their relative winding number. 

We have created a new application of scattering singularities by combining two of them in a novel manner. This combination creates a useful means of generating stable I/Q signals with enhanced tolerance for the properties of the input signals.  

\section*{Acknowledgments}
We acknowledge helpful discussions with Eitan Bachmat. We thank David Shrekenhamer and Timothy Sleasman of JHU APL for the design and fabrication of the metasurfaces used in this work. This work was supported by NSF/RINGS under grant No. ECCS-2148318, ONR under grant N000142312507, DARPA WARDEN under grant HR00112120021.

%End of Main Text

\appendix
\section{LIMITATIONS OF 2 $\times$ 2 SCATTERING MATRIX MODEL}\label{sec.Limitations}
For a general $2\times 2$ scattering matrix, we can write the eigenvalues and eigenvectors as:

\begin{equation}
\lambda_S^{1,2} = \frac{S_{11}+S_{22}}{2} \pm \frac{1}{2}\sqrt{4 S_{12} S_{21} \left( 1 + M_{S}^2 \right)}
\end{equation}

\begin{equation}
 \ket{R_{1,2}} =
\begin{pmatrix}
\frac{1}{S_{21}}\left(\lambda_S^{1,2} - S_{22}\right)\\
1
\end{pmatrix},
\end{equation}
where
\begin{equation}
M_{S} \equiv \frac{S_{11} - S_{22}}{2 \sqrt{S_{12} S_{21}}}.
\label{eqn:MSdef2}
\end{equation}

Note that in the limit that $S_{21}$ goes to 0, the eigenvectors should be of the form $\begin{pmatrix} 1 \\ \frac{1}{S_{12}}\left(\lambda_S^{1,2} - S_{11}\right) \end{pmatrix}$, and in the limit that $S_{21}$ and $S_{12}$ go to 0, the system is trivially diagonalized and has an orthonormal basis. A practical limitation of this model is that $M_{S}$ is not generally continuous due to the complex square root function in its denominator. Approaching ${\cal R}e(\sqrt{S_{12} S_{21}}) = 0$ from the left, yields ${\cal I}m(M_{S}) = X$, but approaching ${\cal R}e(\sqrt{S_{12} S_{21}}) = 0$ from the right, yields ${\cal I}m(M_{S}) = -X$. As parameters of the system vary, the curves of ${\cal R}e(\sqrt{S_{12} S_{21}}) = 0$ move in parameter space, which makes some of the model's predictions difficult to verify, specifically the determination of the winding number and/or charge of the singularities.

There is an ambiguity in determining the charges of EPD's, but not the winding numbers, in the non-reciprocal case. The winding number of the orthogonality points can be determined by looking at the zeros of $i{\cal I}m(M_{S}) + \left| \frac{S_{12}}{S_{21}} \right| - 1$, but this quantity suffers from discontinuities due to the square root in $M_{S}$. There are two charges of EPD's which are the zeros of the complex scalar function $S_{11} - S_{22} \mp 2 i \sqrt{S_{12} S_{21}}$, and only EPD's of the same charge and opposite winding number should be able to pair-create or pair-annihilate. Unfortunately due to the square root discontinuities, information is mixed between the two equations, which makes it difficult to uniquely identify the EPD's. Instead of $M_S$, we can instead examine $M_{S}^2 = \frac{(S_{11} - S_{22})^2}{4 S_{12} S_{21}}$. Here the EPD's correspond to $M_{S}^2 = -1$, but with this function there are only two different EPD's, which are determined by their winding number around zeros of $(S_{11} - S_{22})^2+ 4 S_{12} S_{21}$. For the reciprocal case ($M_S^R$) the winding numbers of all EPD's determined from the zeros of $S_{+i}^R$, $S_{-i}^R$ are the same as those determined from the zeros of $(S_{11} - S_{22})^2+ 4 S_{21}^2$. Therefore while using $(M_S^R)^2$ we lose information about the two charges of EPD's ($\pm i$), but we still retain all of their winding numbers and information about how pairs of EPD's are connected (${\cal R}e(M_S^R) = 0$, $-1 \le {\cal I}m(M_S^R) \le 1$). If two connected EPD's have the same winding number then we know that they must be EPD's of different charge. We believe that this still holds in a non-reciprocal system for $(S_{11} - S_{22})^2+ 4 S_{12} S_{21}$ and ${\cal R}e(M_{S}) = 0$, $-1 \le {\cal I}m(M_S) \le 1$. With a non-reciprocal system we find that some of the EPD's connected have the same winding number and some have the opposite using $(S_{11} - S_{22})^2+ 4 S_{12} S_{21}$. This implies that there are two charges of EPD's ($\pm i$), but we can't unambiguously determine their values from $M_{S}$ due to the square root discontinuities.

\section{JORDAN VECTOR DERIVATION}\label{sec.Jordan}
At an $M_S = \pm i$ exceptional point degeneracy, the eigenbasis is collapsed and there is only a single eigenvector. To span the eigenbasis, we introduce the Jordan vector $\ket{J_{\pm i}}$ which for the $2 \times 2$ scattering matrix is a rank 2 generalized eigenvector. To compute the Jordan vector we solve:

\begin{equation}
(S - \lambda_S^{\pm i} I) \ket{J_{\pm i}} = \ket{R_{\pm i}}, \label{Jvec}
\end{equation}

where $S$ is the scattering matrix, $\lambda_S^{\pm i}$ is the degenerate eigenvalue at the EPD (see Eq. \ref{eval} with $M_S = \pm i$), $I$ is the identity matrix, and $\ket{R_{\pm i}}$ is the degenerate eigenvector (see Eq. \ref{DV}). To simplify the problem, we rewrite $\lambda_S^{\pm i}$ in different forms, utilizing the fact that at an exceptional point degeneracy $M_S = \frac{S_{11} - S_{22}}{2 \sqrt{S_{12} S_{21}}} = \pm i$:

\begin{equation}
\lambda_S^{\pm i} = \frac{S_{11} + S_{22}}{2} = S_{22}\pm i\sqrt{S_{12}S_{21}} = S_{11}\mp i\sqrt{S_{12}S_{21}}. \label{Dif_form}
\end{equation}

Letting $\ket{J_{\pm i}} = \begin{pmatrix} a\\ b \end{pmatrix}$ and using Eq. \ref{Dif_form}, Equation \ref{Jvec} simplifies to

\begin{align}
\begin{split}
\pm i\sqrt{S_{12}S_{21}}a + S_{12}b =  \frac{\pm i \sqrt{{S_{12}}{S_{21}}}}{S_{21}\sqrt{1 + \left| \frac{S_{12}}{S_{21}} \right|}} \\ S_{21}a \mp i\sqrt{S_{12}S_{21}}b =  \frac{1}{\sqrt{1 + \left| \frac{S_{12}}{S_{21}} \right|}}.
\end{split}
\end{align}

Multiplying the bottom equation by $\pm i\frac{\sqrt{S_{12}{S_{21}}}}{S_{21}}$, we see that both equations are the same. Solving the equation for $a$, we find that the Jordan vector has the form

\begin{align}
\begin{split}
\ket{J_{\pm i}} &= \begin{pmatrix} \frac{1}{S_{21}\sqrt{1 + \left| \frac{S_{12}}{S_{21}} \right|}} \pm ib\frac{S_{12}}{\sqrt{S_{12}{S_{21}}}}\\ b \end{pmatrix} \\ 
\ket{J_{\pm i}} &= b \sqrt{1 + \left| \frac{S_{12}}{S_{21}} \right|} \ket{R_{\pm i}} + \frac{1}{S_{21} \sqrt{1 + \left| \frac{S_{12}}{S_{21}} \right|}} \begin{pmatrix} 1\\ 0 \end{pmatrix}, \label{Jordan_vec}
\end{split}
\end{align}

where $b$ is an arbitrary coefficient (which can be determined, for example, by normalization). In general for the degenerate eigenvector with arbitrary normalization, the Jordan vector will have the form

\begin{align}
\begin{split}
\ket{R^{'}} &= c_0 \ket{R} \\ 
\ket{J^{'}} &= c_0 \ket{J} + c_1 \ket{R},
\end{split}
\end{align}

where $c_0 \neq 0,c_1$ are arbitrary \cite{Bronson69,Seyranian03}.

Similar to the limitations of the $2 \times 2$ scattering matrix model explained in Appendix \ref{sec.Limitations}, the form of the Jordan vector above only works if $S_{21} \neq 0$. If $S_{21} = 0$, then we can use the alternate eigenvector formula from Appendix \ref{sec.Limitations} and follow the same derivation above to get the correct Jordan vector. If both $S_{21},S_{12} = 0$, the eigenvectors are orthonormal and EPD's are not possible.

For a reciprocal system ($S_{12} = S_{21}$), following the same derivation as above but replacing $\lambda_S^{\pm i}$ and $\ket{R_{\pm i}}$ with their reciprocal versions (Eqs. \ref{eig_rec}, \ref{evec_rec} with $M_S^R=\pm i$) in Equation \ref{Jordan_vec}, we find the form of the Jordan vector to be:

\begin{align}
\begin{split}
\ket{J_{\pm i}} &= \begin{pmatrix} \frac{1}{S_{21}\sqrt{2}} \pm ib\\ b \end{pmatrix} \\ 
\ket{J_{\pm i}} &= b\sqrt{2} \ket{R_{\pm i}} + \frac{1}{S_{21}\sqrt{2}} \begin{pmatrix} 1 \\ 0 \end{pmatrix}. \label{Jvec_rec}
\end{split}
\end{align}

\section{PROOF OF EIGENVECTOR ORTHOGONALITY CONDITIONS}\label{sec.Proof}

For a general system:

\begin{align}
\braket{R_1|R_2}  = \left( \frac{\lambda_S^{1} - S_{22}}{S_{21}} \right)^*
 \left(\frac{\lambda_S^{2} - S_{22}}{S_{21}} \right) + 1.
\end{align}

\begin{align}
\begin{split}
&\braket{R_1|R_2} = \left(\frac{S_{11}-S_{22}}{2S_{21}} + \frac{1}{2S_{21}}\sqrt{4S_{12}S_{21}(1+M_S^2)} \right)^* \\
& \times \left(\frac{S_{11}-S_{22}}{2S_{21}} - \frac{1}{2S_{21}}\sqrt{4S_{12}S_{21}(1+M_S^2)} \right) + 1
\end{split}
\end{align}

\begin{align}
\begin{split}
&\braket{R_1|R_2} = \left(M_S\frac{\sqrt{S_{12}S_{21}}}{S_{21}} + \frac{1}{2S_{21}}\sqrt{4S_{12}S_{21}(1+M_S^2)} \right)^* \\
& \times \left(M_S\frac{\sqrt{S_{12}S_{21}}}{S_{21}} - \frac{1}{2S_{21}}\sqrt{4S_{12}S_{21}(1+M_S^2)} \right) + 1
\end{split}
\end{align}

\begin{align}
\begin{split}
& \braket{R_1|R_2} = 1+\left|\frac{S_{12}}{S_{21}}M_S^2\right| - \left|\frac{S_{12}}{S_{21}}\left(1+M_S^2\right)\right|\\
& + 2i{\cal I}m\left[ \left( \frac{1}{2S_{21}}\sqrt{4S_{12}S_{21}(1+M_S^2)}\right)^*\left(M_S\frac{\sqrt{S_{12}S_{21}}}{S_{21}}\right) \right]
\end{split}
\end{align}

If ${\cal I}m(M_{S}) = 0$, then

\begin{align}
\begin{split}
& \braket{R_1|R_2} = 1+\left|\frac{S_{12}}{S_{21}}M_S^2\right| - \left|\frac{S_{12}}{S_{21}}\left(1+M_S^2\right)\right|\\
& + 2i{\cal I}m\left[ \left( \frac{1}{S_{21}}\sqrt{S_{12}S_{21}}\sqrt{1+M_S^2}\right)^*\left(M_S\frac{\sqrt{S_{12}S_{21}}}{S_{21}}\right) \right]
\end{split}
\end{align}

\begin{align}
\begin{split}
& \braket{R_1|R_2} = 1+\left|\frac{S_{12}}{S_{21}}M_S^2\right| - \left|\frac{S_{12}}{S_{21}}\left(1+M_S^2\right)\right|\\
& + 2i{\cal I}m\left[ \left(\sqrt{1+M_S^2}\right)^*\left(M_S\left| \frac{S_{12}}{S_{21}}\right| \right) \right]
\end{split}
\end{align}

\begin{align}
\begin{split}
& \braket{R_1|R_2} = 1+\left|\frac{S_{12}}{S_{21}}M_S^2\right| - \left|\frac{S_{12}}{S_{21}}\left(1+M_S^2\right)\right|
\end{split}
\end{align}

\begin{align}
\begin{split}
& \braket{R_1|R_2} = 1+\left|\frac{S_{12}}{S_{21}}\right|\left(M_S^2 - (1+M_S^2)\right)
\end{split}
\end{align}

\begin{align}
\begin{split}
& \braket{R_1|R_2} = 1+\left|\frac{S_{12}}{S_{21}}\right|\left(-1\right)
\end{split}
\end{align}

If $\left| \frac{S_{12}}{S_{21}} \right| = 1$, then

\begin{equation}
\braket{R_1|R_2} = 0.
\end{equation}

Therefore, the points of eigenvector orthogonality must be the same as the points where ${\cal I}m(M_{S}) = 0$ and $\left| \frac{S_{12}}{S_{21}} \right| = 1$.

\section{EPD PAIR WINDING NUMBER-CHARGE RELATION}\label{sec.Phase_Winding}
Every singularity of a complex scalar function in a two-dimensional parameter space can be defined by the intersection of two curves. For EPD's ($\pm i$), two such curves are ${\cal R}e(M_{S}) = 0$ and ${\cal I}m(M_{S}) \mp 1 = 0$. The intersection of these two curves at each $\pm i$ EPD defines four quadrants determined by the signs of ${\cal R}e(M_{S})$ and ${\cal I}m(M_{S}) \mp 1$ respectively. By convention, the phase winds counterclockwise around a singularity in the plane, and with this we can determine the relative winding number of each EPD connected by the finite curve ${\cal R}e(M_{S}) = 0$ and $-1 \le {\cal I}m(M_{S}) \le 1$. 

\begin{figure}[htb]
\centerline{%
\includegraphics[width=8.6cm]{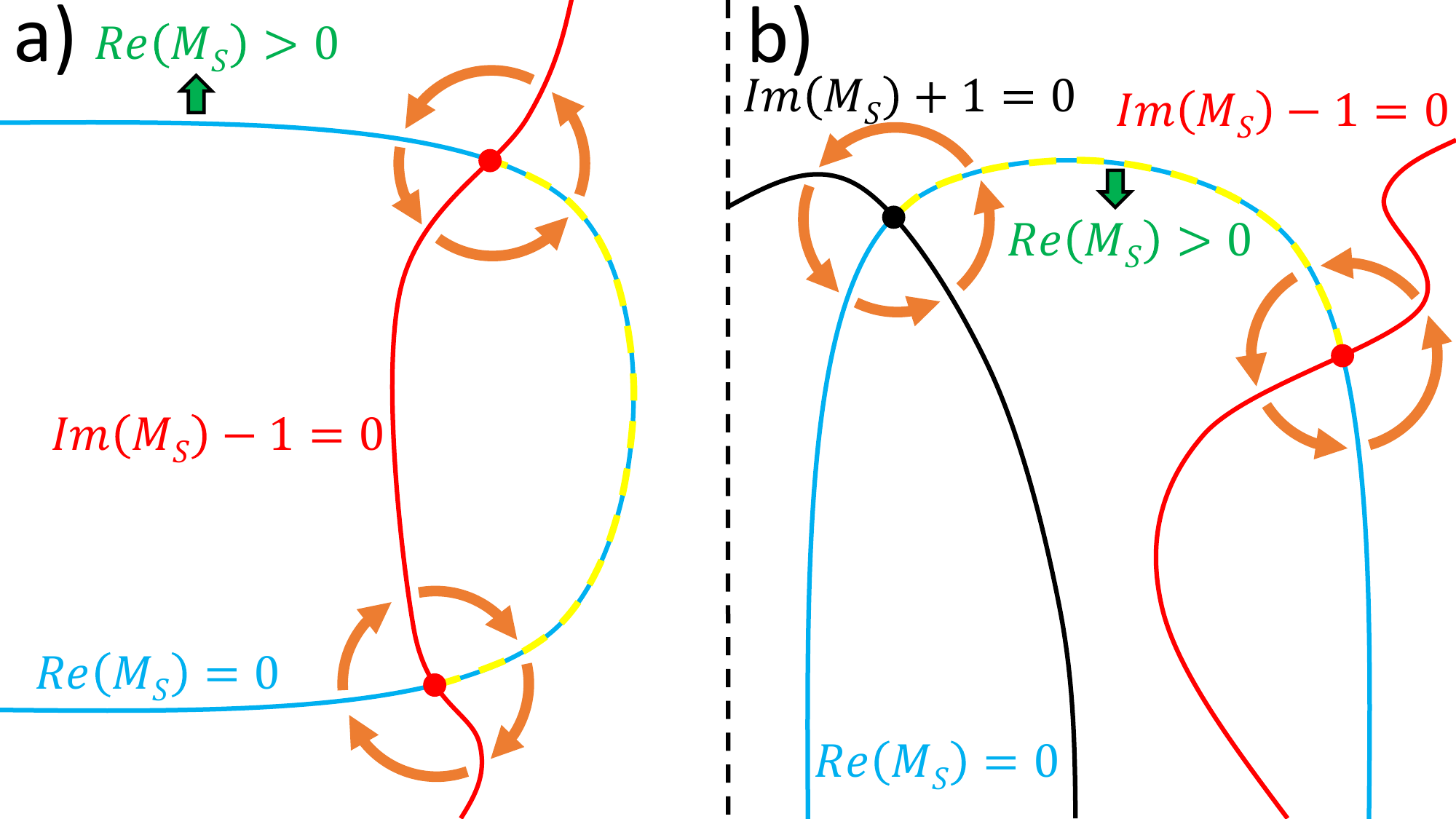}}
\caption{Schematic illustration of phase windings of connected exceptional point degeneracies in an arbitrary two-dimensional parameter space. The yellow dashed curves correspond to ${\cal R}e(M_S) = 0$ and $-1 \le {\cal I}m(M_S) \le 1$ which connects the pairs of EPD's, and the orange curved arrows indicate the direction of the phase winding of each EPD. The red and black dots correspond to the $M_S=+i$ and $M_S=-i$ EPD's, respectively, and the colored equations describe the same-colored curves. The green inequalities denote the arbitrary sign choice of ${\cal R}e(M_S)$. a) Two $+i$ EPD's connected by ${\cal R}e(M_S) = 0$ and $-1 \le {\cal I}m(M_S) \le 1$ that have opposite phase winding. b) A $+i$ and $-i$ EPD connected by ${\cal R}e(M_S) = 0$ and $-1 \le {\cal I}m(M_S) \le 1$ that have the same phase winding. Note that the two other charge/winding number cases are obtained by reversing the arbitrary sign of ${\cal R}e(M_{S})$, which results in reversal of the winding around each EPD.}
\label{EPDs_Winding}
\end{figure}

An illustration of winding number determination for the cases of two equally charged and two oppositely charged EPD's is shown in Figure \ref{EPDs_Winding}. In this figure, we show that EPD pairs of the same charge connected by ${\cal R}e(M_S) = 0$ and $-1 \le {\cal I}m(M_S) \le 1$ have opposite winding numbers, and EPD pairs of opposite charge connected by ${\cal R}e(M_S) = 0$ and $-1 \le {\cal I}m(M_S) \le 1$ have the same winding number.

\section{SIMPLIFICATION OF 2 $\times$ 2 SCATTERING MATRIX MODEL WITH RECIPROCITY}\label{sec.Simplification}
For a $2\times 2$ reciprocal scattering matrix ($S_{12} = S_{21}$), the equations for the eigenvalues, eigenvectors, and $M_{S}$ simplify to:

\begin{equation}
\lambda_S^{1,2} = \frac{S_{11}+S_{22}}{2} \pm \frac{1}{2}\sqrt{4 S_{21}^2 \left( 1 + (M_S^R)^2 \right)} \label{eig_rec}
\end{equation}

\begin{equation}
 \ket{R_{1,2}} =
\begin{pmatrix}
\frac{1}{S_{21}}\left(\lambda_S^{1,2} - S_{22}\right)\\
1
\end{pmatrix}, \label{evec_rec}
\end{equation}

where
\begin{equation}
M_S^R \equiv \frac{S_{11} - S_{22}}{2 S_{21}}
\end{equation}
is a complex scalar function of system parameters, in general. Note that in the limit that $S_{21}$ goes to 0, the system is trivially diagonalized and has an orthonormal basis.

There are many ways to quantify eigenvector overlap in the literature, but generally the Petermann factor or phase rigidity/coalescence parameter are used \cite{Alireza22,Tuxbury2022,Wiersig23}. For a $2\times 2$ matrix, both the Petermann factor $K$ and phase rigidity $r$ can be written in terms of $|C|$:

\begin{equation}
K = \frac{1}{1-\lvert C \rvert^2}
\label{eqn:K}
\end{equation}

\begin{equation}
r = \sqrt{1-\lvert C \rvert^2} \:.
\label{eqn:r}
\end{equation}

From these equations we see that the simplest quantity to measure eigenvector overlap is $|C|$ itself, as $K$ and $r$ are much more nonlinear.

\section{EXPERIMENTAL IMPLEMENTATIONS OF SCATTERING}\label{sec.Exp}
\subsection{Networks of coaxial cables}
Networks of coaxial cables (graphs) are made up of nodes and bonds, along which the wave propagation is described by a one-dimensional (1D) Helmholtz equation. Graphs are a convenient setting for studying wave chaos \cite{Kottos97,Kottos00,Hul04}. Such systems allow the use of simple experimental methods to break reciprocity e.g. by means of microwave circulators on a subset of the nodes of the graph. The graphs have two ports, where the cables leading to the vector network analyzer (VNA) are attached.

We employ a quasi-one-dimensional tetrahedral microwave graph with adjustable phase shifters incorporated into four of the six bonds making up the graph. The phase shifters allow us to parametrically change the lengths of the bonds of the graph. The total electrical length of the reciprocal graph is approximately 3.2 m, and the total electrical length of the non-reciprocal graph is approximately 3.3 m. Each Narda-MITEQ P1507D-SM24 phase shifter can change the phase of the transmitted electromagnetic wave from 0-60 Degrees/GHz over the 0-18.6 GHz frequency range, which is an effective length change of the bond by approximately 5 cm. The phase shifters can be independently controlled with a DAQ card to electronically change the phase shift by a set amount. For the graph with non-reciprocity, we replace one of the internal nodes of the graph with a Narda-MITEQ Model 4925 Circulator, which operates in the frequency band of 7-12.4 GHz.

\subsection{Two-dimensional cavities}
A quasi-two-dimensional rectangular billiard made of brass has two coupling ports on the lid \cite{Gokir98}. The billiard has a height of 7.9 mm and an area of 0.077 $m^2$. When the cavity is excited at frequencies below approximately 19 GHz, only one propagating mode is supported, with the electric field polarized in the vertical (thin) direction \cite{Stock99,So95}. 

The billiard is loaded with three tunable metasurfaces which were fabricated by the Johns Hopkins University Applied Physics Laboratory \cite{PhysRevApplied.20.014004} and designed to vary reflection amplitude between 11-18 GHz and reflection phase between 14-16 GHz. The metasurfaces consist of a linear array of 18 mushroom-shaped resonant elements, where each element is loaded with varactor diodes and is sub-wavelength in size \cite{Siev99}. The metasurfaces were fabricated with Rogers 5880 PCB material and MACOM MAVR-011020-1411 varactor diodes. Each metasurface is 7.9 mm high, 185 mm long, and 1.8 mm thick, and has enough flexibility to conformally attach to a curved interior wall. All diodes on a given metasurface are tuned simultaneously by applying a global DC voltage bias to the metasurface through thin insulated wires that exit the billiard beneath the lid. The capacitance of the varactor diodes decreases from 0.24 to 0.03 pF as the applied voltage bias increases from 0 to 15 V, thus increasing the resonant frequency of the patches. Both the reflection magnitude and phase of the metasurface change as the voltage is varied, in general. Thus the tuned perturbation is a non-Hermitian change to the closed-system Hamiltonian $\mathcal{H}$ and the spectrum of the operator (modes of the cavity). The metasurfaces are connected to a Keithley 2230G-30-1 triple channel programmable DC power supply and are positioned along three different walls of the billiard. Each metasurface covers approximately 16\% of the perimeter length of the billiard. Through the variation of the reflection coefficient of the metasurfaces, we can manipulate the Hamiltonian, and therefore the scattering matrix, to create conditions for exceptional point degeneracies, coherent perfect absorption \cite{Erb24}, and other types of scattering singularities. Note that the total length of the metasurface is greater than the microwave wavelength, making them intermediate in character between a purely local perturbation and a global perturbation (such as the driving frequency). 

\subsection{Three-dimensional cavities}
A nearly-cubic rectangular three-dimensional cavity is connected to the outside world through two ports. The cavity has side lengths of approximately 0.92 m and a volume of $\sim$0.76 $m^3$ \cite{Frazier20}. Inside the cavity are two two-dimensional metasurfaces and various irregularly-shaped scatterers to increase the system's complexity. The two-dimensional metasurfaces used in this cavity were fabricated by the Johns Hopkins University Applied Physics Laboratory \cite{PhysRevApplied.20.014004} and were designed to vary reflection amplitude between 2-3.6 GHz and reflection phase between 3-3.5 GHz. The metasurfaces are made up of a 2D square array of $14\times 14$ mushroom-shaped resonant elements, where each element is loaded with varactor diodes and is sub-wavelength in size \cite{Siev99}. The metasurfaces were fabricated with Rogers 4003c PCB material and Skyworks SMV1405-079LF varactor diodes. Each metasurface is 26 cm by 26 cm in size, and takes up approximately 1.3\% of the interior surface area of the enclosure. The varactor diodes on a given metasurface can be tuned simultaneously with a globally applied DC voltage bias to the metasurface through shielded cables that exit the cavity through the top wall. As the applied voltage bias is increased from 0 to 30 V, the capacitance of the varactor diodes varies, thus increasing the resonant frequency of the patches. Both the reflection magnitude and phase of the metasurface change as the voltage is varied, in general. Thus the tuned metasurface presents a non-Hermitian modification to the closed-system Hamiltonian, changing its spectrum (i.e. the modes of the cavity). The metasurfaces are placed along two different walls of the cavity and are connected to a Keithley 2230G-30-1 triple channel programmable DC power supply. Note that the area of the metasurface is greater than the square of the microwave wavelength, making them intermediate in character between a purely local perturbation and a global perturbation. 

Despite the presence of varactor diodes on the one-dimensional and two-dimensional metasurfaces, we operate the cavities in the low-power linear-response limit. Under these conditions the scattering properties of the system are fully captured by the linear scattering matrix.

\section{EIGENVECTOR COALESCENCE OF RECIPROCAL SCATTERING SYSTEMS}\label{sec.Reciprocal_Systems}
\begin{figure}[htb]
\centerline{%
\includegraphics[width=9.1cm]{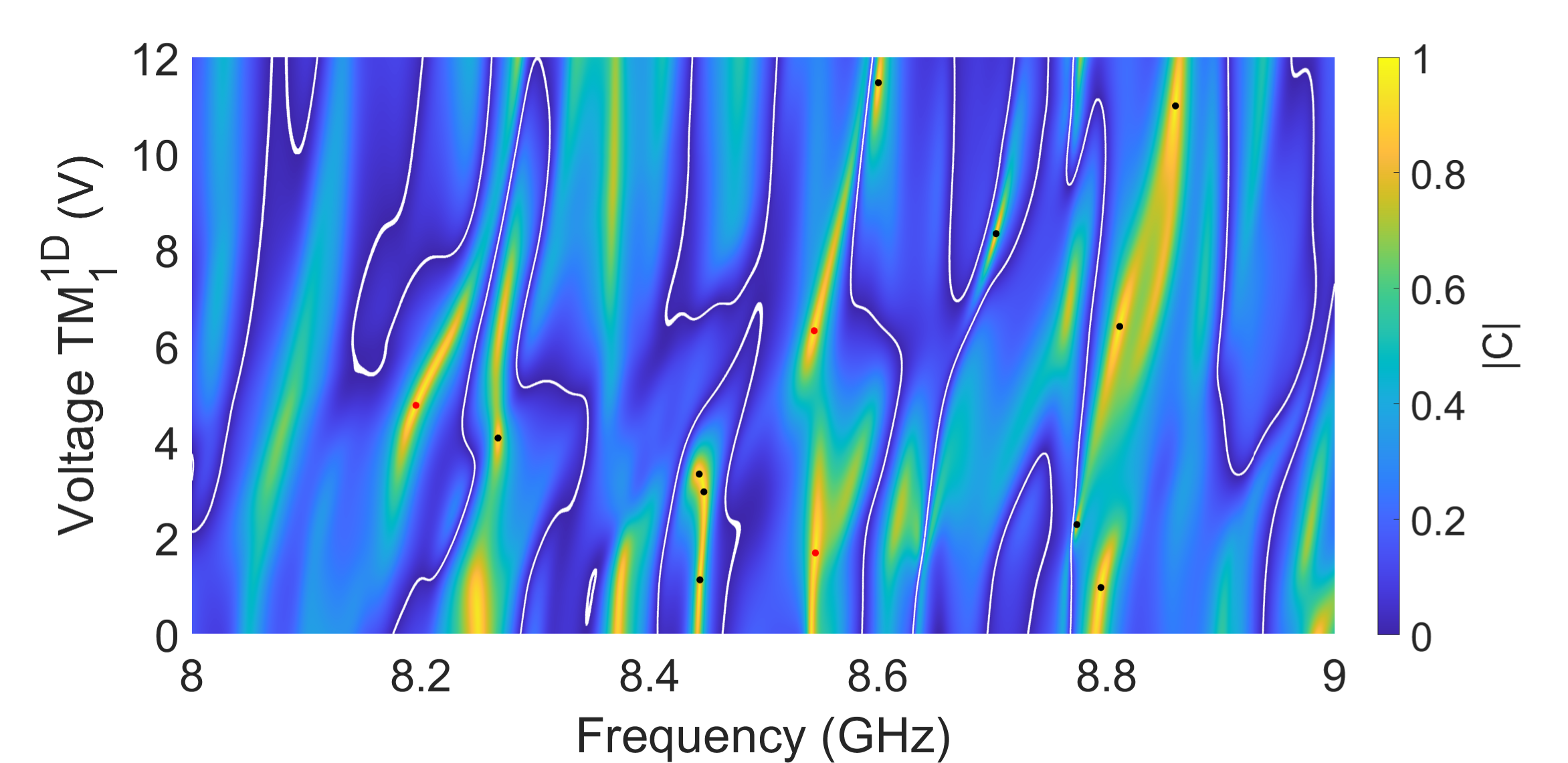}}
\caption{Experimental exceptional point degeneracies and orthogonality curves in two-dimensional parameter space. The red and black dots correspond to the $M_S^R=+i$ and $M_S^R=-i$ EPD's respectively. The white regions are points of near zero eigenvector coalescence ($|C|<0.002$). Eigenvector coalescence $|C|$ vs frequency and $TM_1^{1D}$ applied bias voltage for the ray-chaotic quarter bow-tie billiard \cite{So95,Gokir98,Erb24}.}
\label{Quarter_Bowtie_Coalescence}
\end{figure}

\begin{figure}[htb]
\hspace*{-0.2cm}
\centerline{%
\includegraphics[width=8.7cm]{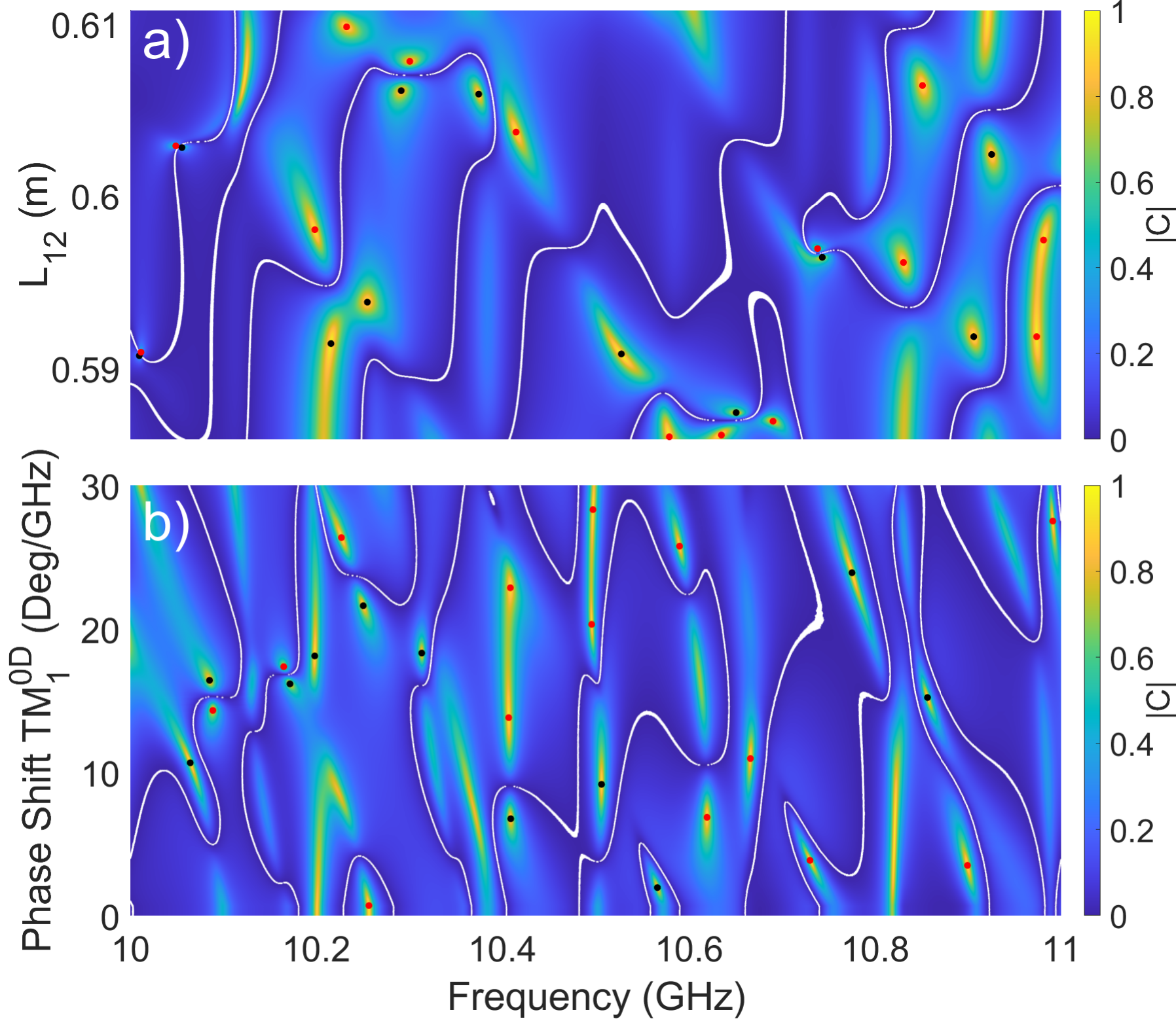}}
\caption{Exceptional point degeneracies and orthogonality curves in two-dimensional parameter spaces. The red and black dots correspond to the $M_S^R=+i$ and $M_S^R=-i$ EPD's respectively. The white regions are points of near zero eigenvector coalescence ($|C|<0.002$). a) Eigenvector coalescence $|C|$ vs frequency and bond length $L_{12}$ for the analytic model of a tetrahedral graph. b) Eigenvector coalescence $|C|$ vs frequency and phase shift of $TM_1^{0D}$ the experimental tetrahedral graph.}
\label{Recip_Graph}
\end{figure}

\begin{figure}[htb]
\centerline{%
\includegraphics[width=9.1cm]{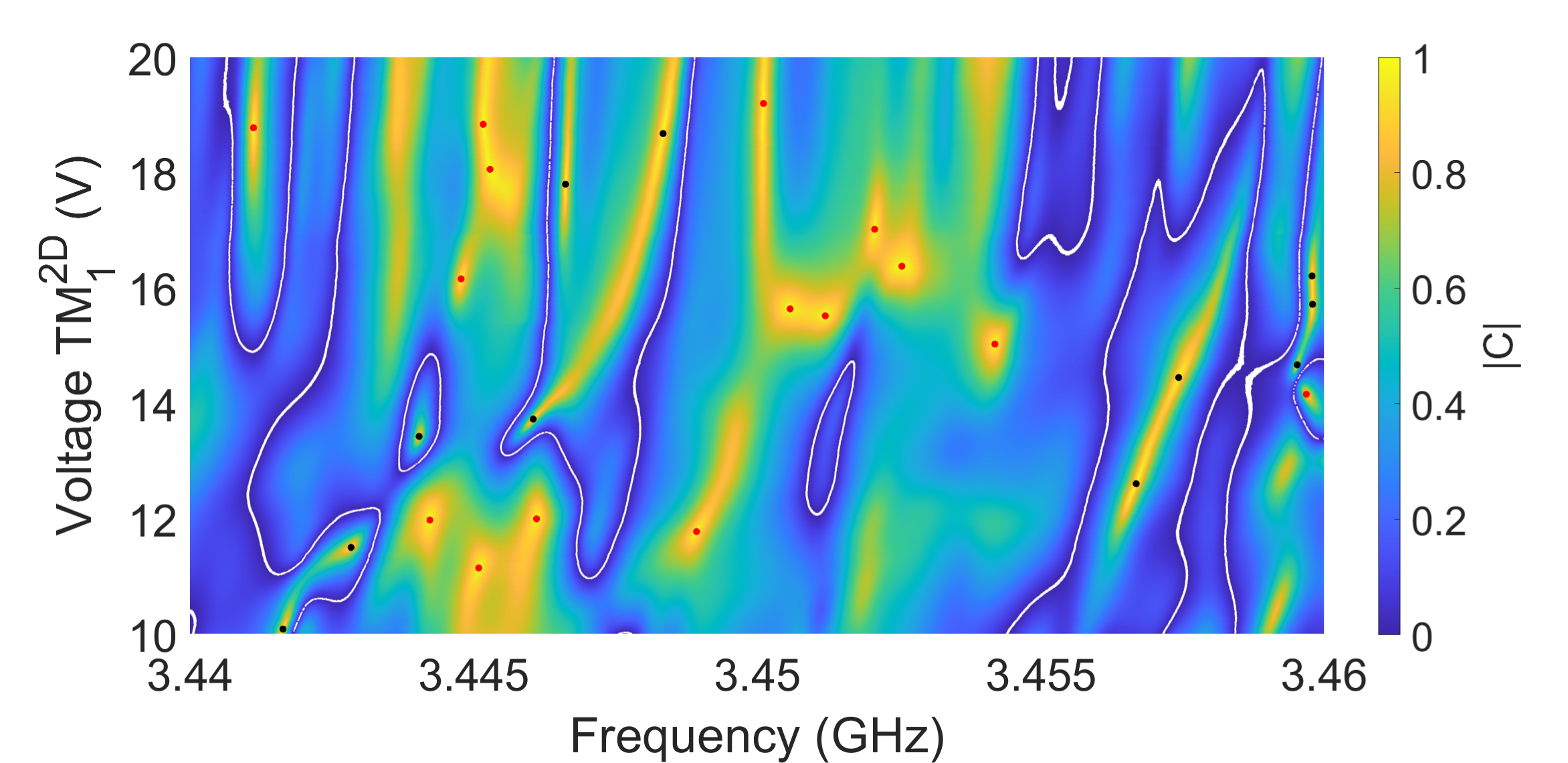}}
\caption{Experimental exceptional point degeneracies and orthogonality curves in two-dimensional parameter space. The red and black dots correspond to the $M_S^R=+i$ and $M_S^R=-i$ EPD's respectively. The white regions are points of near zero eigenvector coalescence ($|C|<0.004$). Eigenvector coalescence $|C|$ vs frequency and $TM_1^{2D}$ applied bias voltage for the three-dimensional microwave cavity.}
\label{3D_Coalescence}
\end{figure}

All reciprocal systems measured in this work have a structurally similar eigenvector coalescence $|C|$ in two-dimensional parameter spaces. There are curves of eigenvector orthogonality that separate the two different types of exceptional point degeneracies. In the main text, we showed examples of $|C|$ vs two tunable parameters for an experimental rectangular billiard and for a Random Matrix Theory model (Figure \ref{Jul12_Coalescence}a,c). Here we show that the same eigenvector coalescence structure applies for a ray-chaotic quarter bow-tie billiard (Fig. \ref{Quarter_Bowtie_Coalescence}), an experimental tetrahedral graph and the analytic model of a tetrahedral graph (Fig. \ref{Recip_Graph}), and a chaotic three-dimensional microwave cavity (Fig. \ref{3D_Coalescence}).

\section{RANDOM MATRIX THEORY MODELING}\label{sec.RMT}
A typical theoretical modeling of wave scattering invokes a temporal coupled mode theory formulation. In the frequency domain, this modeling is separated in two parts:

\begin{equation}
\omega \ket{\psi} = H_{eff} \ket{\psi} + i W \ket{\alpha_{in}}
\end{equation}

\begin{equation}
\ket{\alpha_{out}} = W^T \ket{\psi} - \ket{\alpha_{in}}
\end{equation}
where the first equation describes the steady-state field $|\psi\rangle$ in the scattering domain, while the second equation describes the input-output relation between the injected monochromatic field $|\alpha_{in}\rangle$ and the outgoing field $|\alpha_{out}\rangle$. The frequency $\omega$ is the frequency of the injected monochromatic wave, $H_{eff}$ is the effective Hamiltonian given by $H_{eff} = H_0 - \frac{i}{2} W^T W$ that describes the scattering domain -- including its coupling to the interrogating channels. This coupling is described by the $2\times N$ coupling matrix $W$ while $H_0$ is the effective Hamiltonian that describes the isolated system (decoupled from the interrogating channels). The scattering matrix $S$ that connects incoming with outgoing waves $|\alpha_{out}\rangle=S|\alpha_{in}\rangle$, takes the form $S(\omega) = -1_2 + i W G(\omega)W^T$, where $G(\omega) = \frac{1}{\omega - H_{eff}}$.

When the scattering domain is a cavity that supports underlying chaotic ray dynamics, the Hamiltonian $H_0$ is modeled by a random matrix ensemble of specific symmetry class. In case of time-reversal symmetric (TRS) systems this ensemble is the Gaussian Orthogonal Ensemble (GOE), while when TRS is violated the appropriate ensemble is the Gaussian Unitary ensemble (GUE) \cite{Stock99,Haake10}. To model a system with additional degrees of freedom, such as metasurfaces, we have introduced an ensemble of random matrices $H_0$ that depends on two parameters $x,y$ and takes the form:

\begin{equation}
H_0(x,y) = H_1 + \lvert cos(x) \rvert \lvert cos(y) \rvert H_2 + \lvert sin(x) \rvert \lvert sin(y) \rvert H_3.
\label{eqn:H0Def2}
\end{equation}

The elements of $H_1$ are taken from a Gaussian Orthogonal Ensemble (GOE) of mean zero and standard deviation $\sqrt{\frac{N}{\pi}}$. Similarly, the off-diagonal elements of $H_2$ and $H_3$ are taken from a GOE of the same family as $H_1$, while the diagonal elements are taken from the uniform distribution $[0,-i]$. In the framework of our experimental setups (see above), the Hamiltonian $H_1$ models the unperturbed chaotic cavities while the non-Hermitian Hamiltonians $H_{2,3}$ model the metasurfaces that are used to perturb the cavities. The trigonometric functions present in Eq. (\ref{eqn:H0Def2}) serve to perturb the system such that the norm of the Hamiltonian is conserved. The added degrees of freedom of $H_0$ make the scattering matrix a function of $x$, $y$, and $\omega$, which allow us to better understand the scattering properties of such systems.

To explore the impact of non-reciprocity on the system, we introduce a strength-controlled magnetic field:
\begin{equation}
H_{mag} = H_0 + i \alpha B,
\end{equation}
where $B$ is an anti-symmetric GOE matrix from the same family as $H_0$ and $\alpha$ is the parameter that controls the relative strength of the magnetic field. With the parameters of $H_0$, $\omega$, and the inclusion of the magnetic field, we can thoroughly explore the dynamics and parametric dependence of the scattering matrix and its singularities for any arbitrary system with or without reciprocity.

\section{THEORETICAL MODELING OF COAXIAL NETWORKS}\label{sec.Networks}

The theoretical modeling of the coaxial network follows Ref. \cite{Kottos03,Wang24}. We consider a microwave network with $\nu=1,\cdots, N$ vertices, two of which are connected to transmission lines (TL) $\beta=1,2$. Two vertices $\nu,\mu$ may be connected by an edge (bond) of length $L_{\nu\mu}$. We define the position along a bond as $x_{\nu\mu}=x$ with $x_{\nu\mu}=0$ and $x_{\nu\mu} = L_{\nu\mu}$ on vertices $\nu$ and $\mu$ respectively. The wave $\psi_{\nu\mu}(x)$ on the $(\nu,\mu)-$bond is a solution of the Helmholtz equation:

\begin{equation}
\frac{d^2}{dx^2}\psi_{\nu\mu}(x) + k^2\psi_{\nu\mu}(x) = 0,
\label{eqn:waveDef}
\end{equation}
where $k=n_{\nu\mu}^{(r)} 2\pi f/c_0$ the wave number, $f$ is the linear frequency of the injected monochromatic wave, $c_0$ the speed of light, and $n_{\nu\mu}^{(r)}$ is the complex-valued relative index of refraction that includes the losses of the $(\nu,\mu)$-coaxial cable. Since all coaxial cables used in our network systems are made from the same materials, $n_{\nu\mu}^{(r)} = n^{(r)}$.

The wave function on the bonds of the network can be written as $\psi_{\nu\mu}(x) = \xi_\nu \frac{sin \: k(L_{\nu\mu} - x)}{sin \: k L_{\nu\mu}} + \xi_\mu \frac{sin \: k x}{sin \: k L_{\nu\mu}}$ where $\psi_{\nu\mu}(0)=\xi_\nu$ and $\psi_{\nu\mu}(L_{\nu\mu})=\xi_\mu$ are the values of the field at the vertices $\Xi=(\xi_1,\cdots,\xi_N)^T$. At the $\beta$-TL (attached to the $\nu$-th vertex) the field takes the form $\psi_\nu^{(\beta)}(x) = I_{\beta} e^{-ikx} + O_{\beta} e^{ikx}$  for $x\geq 0$ where $x=0$ is the position of the $\nu-$th vertex and the coefficients $I_{\beta}, O_{\beta}$ indicate the amplitude of the incoming and outgoing waves on the $\beta$-TL respectively. On each vertex $\nu$, the continuity of the field and the current conservation have to be satisfied. These conditions allow us to express the field amplitudes $\Xi$ at the vertices as:
\begin{equation}
\Xi = 2i \frac{1}{h + i W^T W} W^T I
\label{Xi}
\end{equation}
where the two-dimensional vector $I$ contains the amplitudes $I_\beta$ of the incident field and $W$ is the $2\times N$ matrix describing the coupling between the TL and the vertices. The matrix elements of $W$ take the values $1$ if the $\beta$-TL is attached to the vertex $\nu$ and $0$ otherwise. The $N\times N$ matrix $h$ takes the form

\begin{equation}
h_{\nu \mu}(k) = 
\begin{cases}
-\sum_{l\neq \nu} {\cal A}_{\nu l}cot(k L_{\nu l}), & \nu=\mu \\
{\cal A}_{\nu \mu}csc(k L_{\nu \mu})\cdot e^{i \phi_{\nu \mu}}, & \nu\neq \mu
\end{cases}
\label{MDefh}
\end{equation}

where ${\cal A}$ is the adjacency matrix having elements ${\cal A}_{\nu\mu}=1$ if the vertices $\nu,\mu$ are connected via a bond and zero otherwise.

Finally, by invoking the continuity condition at the vertices where the TL are attached, together with the expression for the field $\Xi$ (see Eq. (\ref{Xi})), the scattering matrix $S$ can be derived. The latter takes the form
\begin{equation}
S = -1_{2} + 2i W \frac{1}{h + i W^T W} W^T.
\end{equation}

Finally, a magnetic field can be effectively introduced via a vector potential along the bonds of the network \cite{Kottos03}, and can be incorporated into the modeling by a phase factor $e^{i\phi_{\nu\mu}}$ (with $\phi_{\nu\mu}=-\phi_{\mu\nu}$) which multiplies the off-diagonal terms of $h$, see Eq. (\ref{MDefh}).

\section{DEMONSTRATION OF SINGULARITIES USING A SIMPLE NETWORK}\label{sec.Simple_Network}

\begin{figure*}[htb]
\centerline{%
\includegraphics[width=18cm]{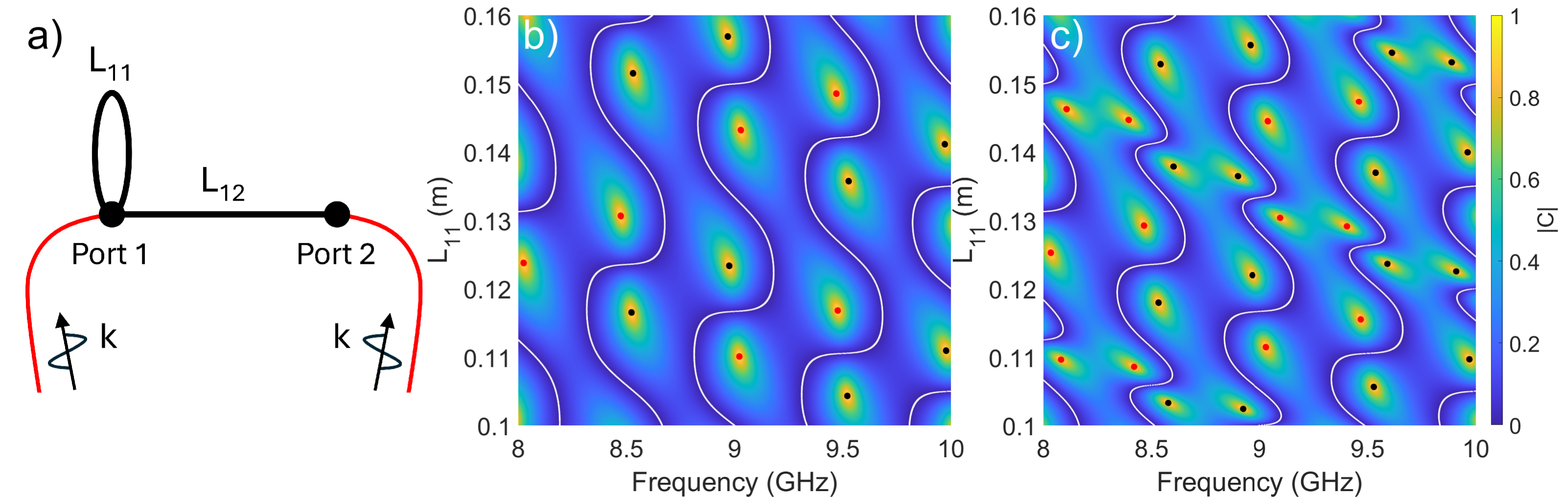}}
\caption{Simple graph schematic, exceptional point degeneracies, and orthogonality curves in two-dimensional parameter spaces. The red and black dots in b), c) correspond to the $M_S=+i$ and $M_S=-i$ EPD's respectively. The white regions are points of near zero eigenvector coalescence ($|C|<0.002$). a) Schematic view of the simple graph. b) Eigenvector coalescence $|C|$ vs frequency and bond length $L_{11}$ for the analytic model of the simple graph with no magnetic field present ($\phi=0$). c) Eigenvector coalescence $|C|$ vs frequency and bond length $L_{11}$ for the analytic model of the simple graph with a magnetic field present ($\phi \neq 0$).}
\label{Simple_Graph}
\end{figure*}

For analytical purposes, we consider a very simple graph consisting of only two vertices and two bonds, one between the two vertices and a loop bond of length $L_{11}$ attached to the first vertex (see Fig. \ref{Simple_Graph}a). In general, one can assume that the loop is threaded by a flux $\phi$. In the case where loops are attached to a vertex $\nu$, the general expression for the matrix $h$ in Eq. (\ref{eqn:MDef}) needs to be modified by adding an additional term $h_{\nu\mu}\rightarrow h_{\nu\mu}+\delta_{\nu\mu}h_{\nu\nu}^{(\rm loop)}$ where $h_{\nu\nu}^{\rm(loop)}=-2\left[\cos(kL_{\nu\nu})-\cos(\phi_\nu)\right]/\sin(kL_{\nu\nu})$.

The corresponding scattering matrix associated with the graph of Fig. \ref{Simple_Graph}a where the loop is coupled to the vertex $\nu=1$, takes the form,

\begin{align}
\begin{split}
S(k) & = \frac{i}{(1 + icot(kL_{12}))(2 - ih_{11}^{\rm(loop)})} \\
& \:\:\: \times \begin{pmatrix}
\rho_+ & 2csc(kL_{12}) \\
2csc(kL_{12}) & -\rho_-
\end{pmatrix},
\label{eqn:S_loop}
\end{split}
\end{align}

where $\rho_{\pm} = h_{11}^{\rm(loop)}(1 \pm icot(kL_{12}))$. Finally, with Eq. (\ref{eqn:S_loop}) we can evaluate the complex scalar function $M_S \equiv \frac{S_{11} - S_{22}}{2 \sqrt{S_{12} S_{21}}}$ as: 

\begin{equation}
M_S = \frac{1}{2} h_{11}^{\rm(loop)} sin(kL_{12}).
\end{equation}

Due to the simplicity of this graph $S_{12} = S_{21}$ always, even in the presence of a magnetic field. Therefore there are always curves of eigenvector orthogonality which will partition the parameter space. In Fig. \ref{Simple_Graph}b we report the coalescence $|C|$ vs frequency and bond length $L_{11}$ in the absence of a magnetic field ($\phi=0$). In Fig. \ref{Simple_Graph}c, we plot the same data from Fig. \ref{Simple_Graph}b, but now in the presence of a magnetic field ($\phi \neq 0$).

\section{EPD PAIR CREATION AND ANNIHILATION}\label{sec.EPD_Creation}
In Figure \ref{Aug16_Movie}, we showed a detailed view of the creation and annihilation of two pairs of exceptional point degeneracies over four settings of the third parameter of the system. In Video \ref{vid.EP_Creation}, we show a much broader view of the evolution of the EPD's over a large sweep of the third parameter. In the video, the winding number of each EPD isn't shown, but they are the same as indicated in Fig. \ref{Aug16_Movie}.

\begin{video}
\hspace*{-0.3cm}
\includegraphics[width=8.9cm]{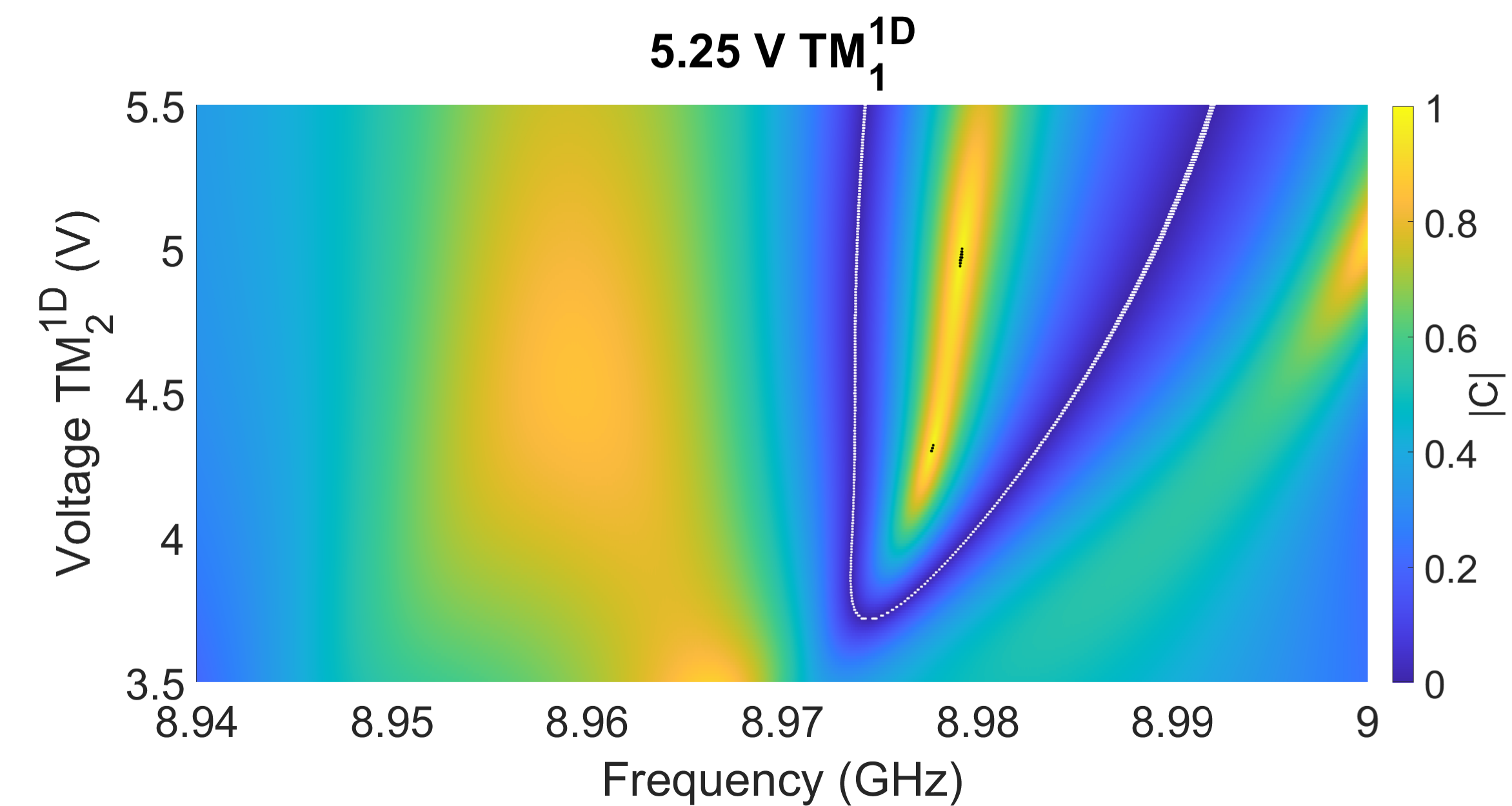}
%\setfloatlink{http://some.video.com/fun.mov}
\caption{\label{vid.EP_Creation} Exceptional point degeneracy creation and annihilation events. The image above shows the first frame of a video which explores a larger parameter space than shown in Figure \ref{Aug16_Movie}. Each frame of the video is at a different fixed applied bias voltage of $TM_1^{1D}$. In the video the $+i$ EPD's are highlighted by the red points where $|{\cal R}e(M_S^R)| \le 0.005$ and $|{\cal I}m(M_S^R) - 1| \le 0.005$, the $-i$ EPD's are highlighted by the black points where $|{\cal R}e(M_S^R)| \le 0.01$ and $|{\cal I}m(M_S^R) + 1| \le 0.01$, and the white eigenvector orthogonality curves are marked by $|C| \le 0.005$.}
\end{video}

\section{EPD DOMAINS AND CONNECTION CURVES}\label{sec.Domains}
\begin{figure*}[htb]
\hspace*{-0.4cm}
\centerline{%
\includegraphics[width=18.2cm]{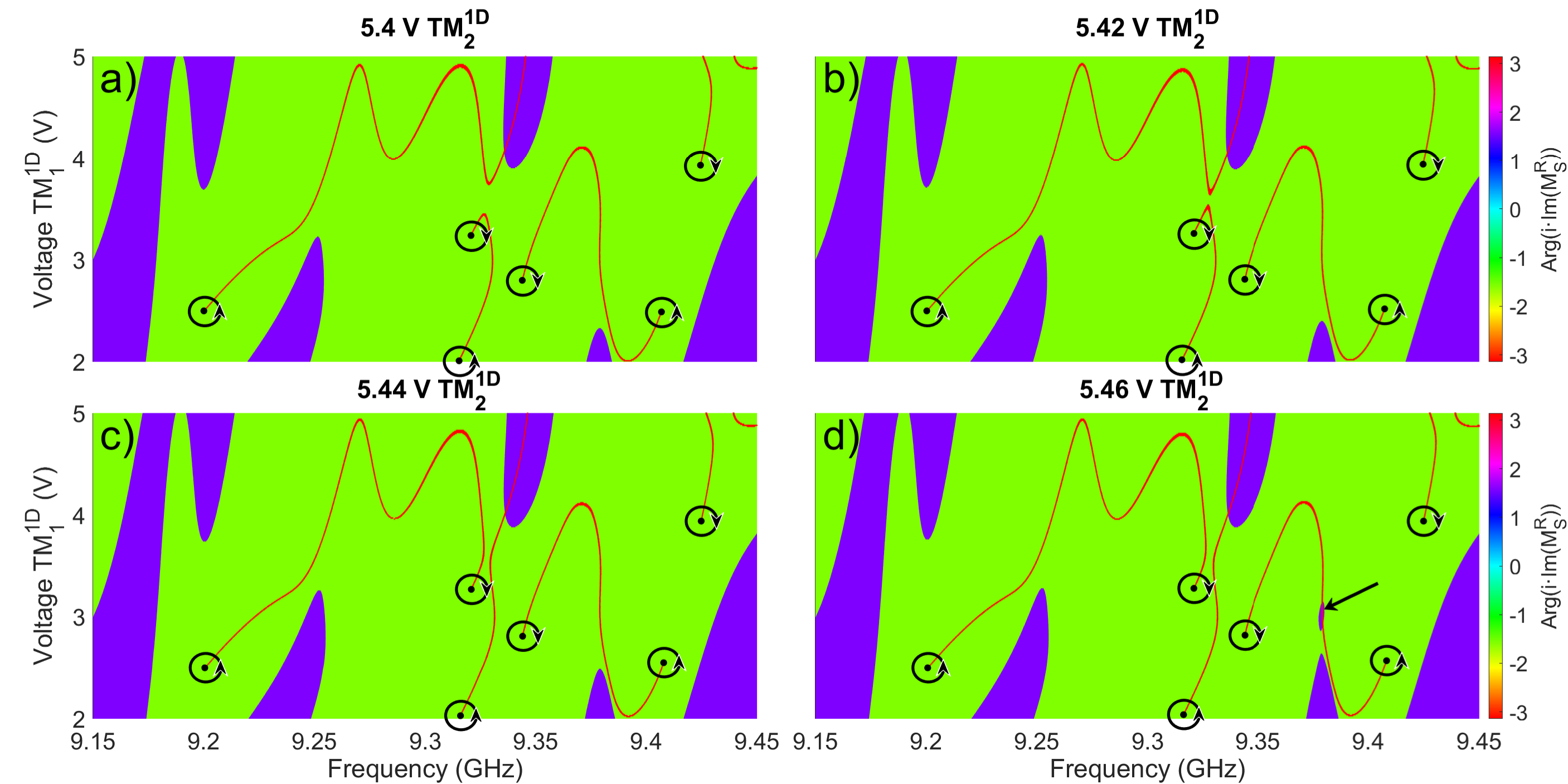}}
\caption{Illustration of reconnection event between curves connecting two pairs of EPD's in a reciprocal system. Plot of $Arg(i\cdot {\cal I}m(M_S^R))$ vs frequency and $TM_1^{1D}$ applied bias voltage in the experimental rectangular billiard. Each of the four plots (a-d) are at a different fixed applied bias voltage of $TM_2^{1D}$. A reconnection event occurs near the center of the panel between b and c. In d, the black arrow indicates the creation of a $+i$ EPD domain inside a $-i$ EPD domain. The domains that the two types of EPD's are allowed to live in are indicated by the green and purple regions. The green regions have a phase of $-\pi /2$ ($-i$ EPD's can live here) and the purple regions have a phase of $+\pi /2$ ($+i$ EPD's can live here). The black dots mark the $M_S^R=-i$ exceptional point degeneracies and the red curves correspond to points of ${\cal R}e(M_S^R) = 0$ and $-1 \le {\cal I}m(M_S^R) \le 1$. }
\label{Aug20_Domain_Movie}
\end{figure*}

Figure \ref{Aug20_Domain_Movie} shows a detailed view of a section of 4 frames from Video \ref{vid.Domains} to highlight the winding numbers of EPD's, the switching of partners of two connected EPD pairs, and the creation of an EPD domain in Fig. \ref{Aug20_Domain_Movie}d. In this figure, the pairs of EPD's connected by ${\cal R}e(M_S^R) = 0$ and $-1 \le {\cal I}m(M_S^R) \le 1$ (red curves) fully within view are all of the same charge, but each EPD in a pair has opposite winding number. As the third parameter of the system varies, two EPD connection curves intersect and exchange partners. This can be seen near the center of the panel between Figures \ref{Aug20_Domain_Movie}b,c. Even though partners were exchanged, the connected EPD's still follow the charge/winding number relation specified in Section \ref{Sec:SandEPD}/Appendix \ref{sec.Phase_Winding}.

\section{EPD CONNECTION CURVES AND THEIR EVOLUTION IN NON-RECIPROCAL SYSTEMS}\label{sec.NR_EPD_Connections}
For non-reciprocal systems, all EPD's are connected in pairs by curves of ${\cal R}e(M_S) = 0$ and $-1 \le {\cal I}m(M_S) \le 1$, although we are unable to resolve their charges (see Appendix \ref{sec.Limitations}). In Video \ref{vid.Non_Reciprocal_Connections}, we show the exact same results as in Video \ref{vid.Non_Reciprocal}, but now we have added the EPD connection curves in red in order to view their evolution as the third parameter of the system is varied. Similar to the reciprocal case, the EPD connection curves can switch partners as the system evolves. When an EPD pair is created (annihilated) the connection curve between two EPD's can be split (connected), as seen around 10.65 GHz near the frame where $TM_2^{0D}= 36.45$ ($TM_2^{0D}= 36.05$).

\begin{video}
\hspace*{-0.44cm}
\includegraphics[width=8.9cm]{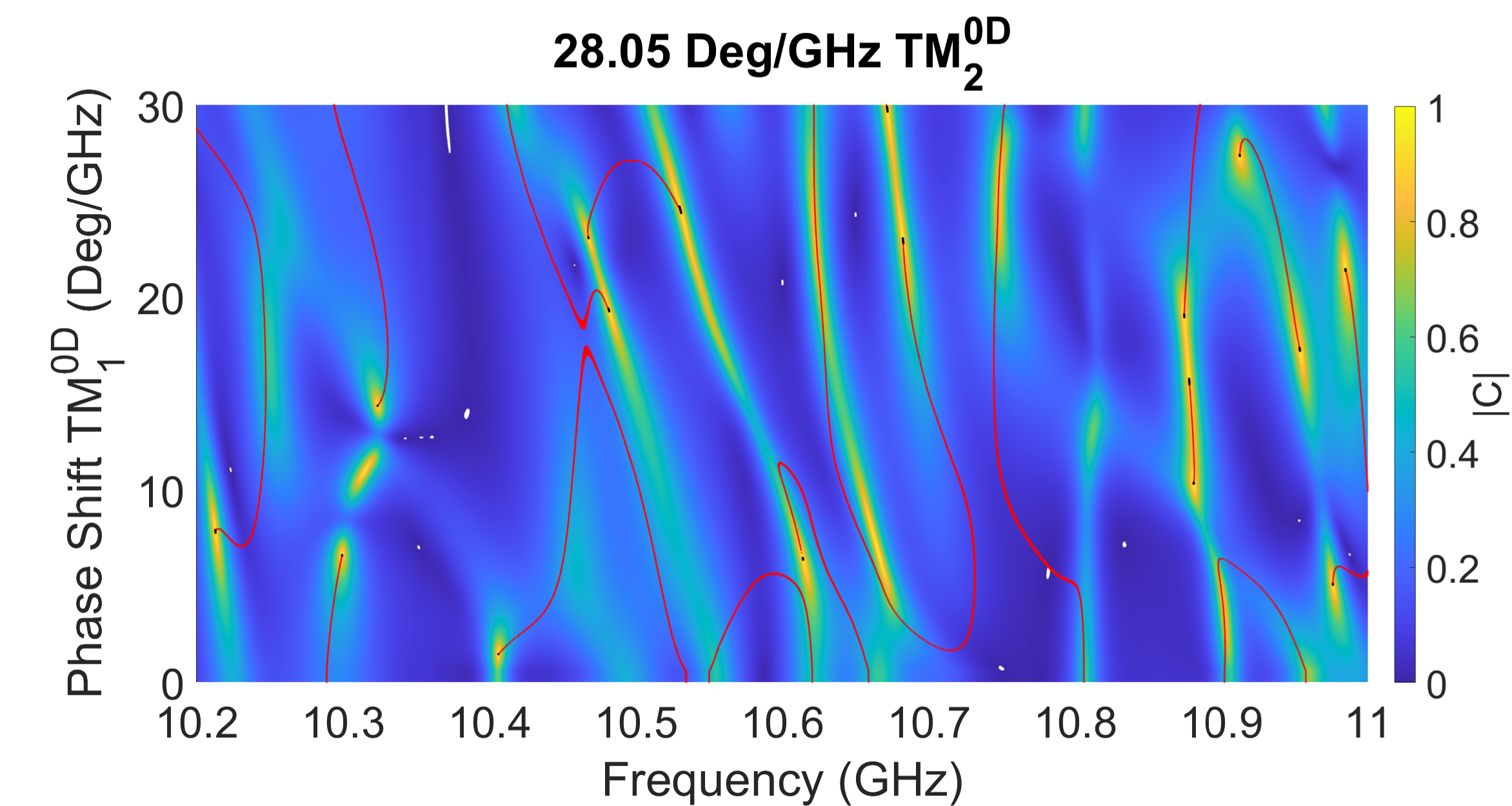}
%\setfloatlink{http://some.video.com/fun.mov}
\caption{\label{vid.Non_Reciprocal_Connections} Dynamics and interactions of exceptional point degeneracies, orthogonality points, and EPD connection curves as a third parameter of the system is varied. The data shown in this figure is exactly the same as shown in Video \ref{vid.Non_Reciprocal}, but with the EPD connections curves included. The image above shows the eigenvector coalescence $|C|$ vs frequency and $TM_1^{0D}$ phase shift for the experimental tetrahedral graph with non-reciprocity for the first frame of the video. Each frame of the video is at a different fixed phase shift of $TM_2^{0D}$. In the video the EPD's are highlighted by the black points where $|C| \ge 0.985$, the white eigenvector orthogonality points are marked by $|C| \le 0.002$, and the red EPD connection curves are marked by $|{\cal R}e(M_S^R)| \le 0.003$ and $|{\cal I}m(M_S^R)| \le 1$. In the video we see numerous EPD/OP creation and annihilation events and the reconnection events of the EPD connection curves between different EPD pairs as the phase shift of $TM_2^{0D}$ varies.}
\end{video}

\section{CPA+EPD EXPERIMENTAL VERIFICATION PROCEDURE}\label{sec.Verification}
In order to first find a location in parameter space roughly near a CPA+EPD condition, we measure the scattering matrix vs frequency for many settings of the other tunable parameters. During these measurements and the subsequent ones, the system is uncalibrated. Next we analyze the scattering eigenvalues of the entire parameter space measured and choose the particular parameter settings that had the smallest value of $|\lambda_S^1| + |\lambda_S^2|$. To verify that this is a good candidate, we also check if the minimum of $|det(S)|$ (CPA condition) and the maximum of $|C|$ (EPD condition) are roughly aligned at the candidate frequency. From there, we perform fine sweeps of each tunable parameter until we find the optimal overall settings such that at a particular point both scattering eigenvalues are very near $\lambda_S = 0+i0$. Then we switch the microwave VNA to the 2-port dual-source mode, so that we can inject an arbitrary signal into the system using both ports simultaneously. To verify the robustness of the CPA+EPD, we inject the arbitrary signals at the CPA+EPD condition over a wide range of relative amplitudes and phases to demonstrate that the output signal power ratio and relative phase between both ports are roughly constant. 

When using the 2-port dual-source mode of the VNA, the input and output signals are: 

\begin{equation}
\ket{\alpha_{\rm in}} = \begin{pmatrix} V_1^{\rm in} e^{i \theta_1}\\ V_2^{\rm in} e^{i \theta_2}\end{pmatrix}, \: \ket{\alpha_{\rm out}} = \begin{pmatrix} V_1^{\rm out} e^{i \phi_1}\\ V_2^{\rm out} e^{i \phi_2}\end{pmatrix}.
\end{equation}

When injecting the arbitrary signals into the system, we measure the input and output power ratios between both ports and their phase difference. The input (output) phase difference is $\theta_1 - \theta_2$ ($\phi_1- \phi_2$). The input and output power ratios between both ports are:

\begin{equation}
\frac{P_1^{in}}{P_2^{in}} = \frac{\left| V_1^{in} \right|^2}{\left| V_2^{in} \right|^2} ,\:\: \frac{P_1^{out}}{P_2^{out}} = \frac{\left| V_1^{out} \right|^2}{\left| V_2^{out} \right|^2}.
\end{equation}

\begin{figure}[htb]
\centerline{%
\includegraphics[width=8.7cm]{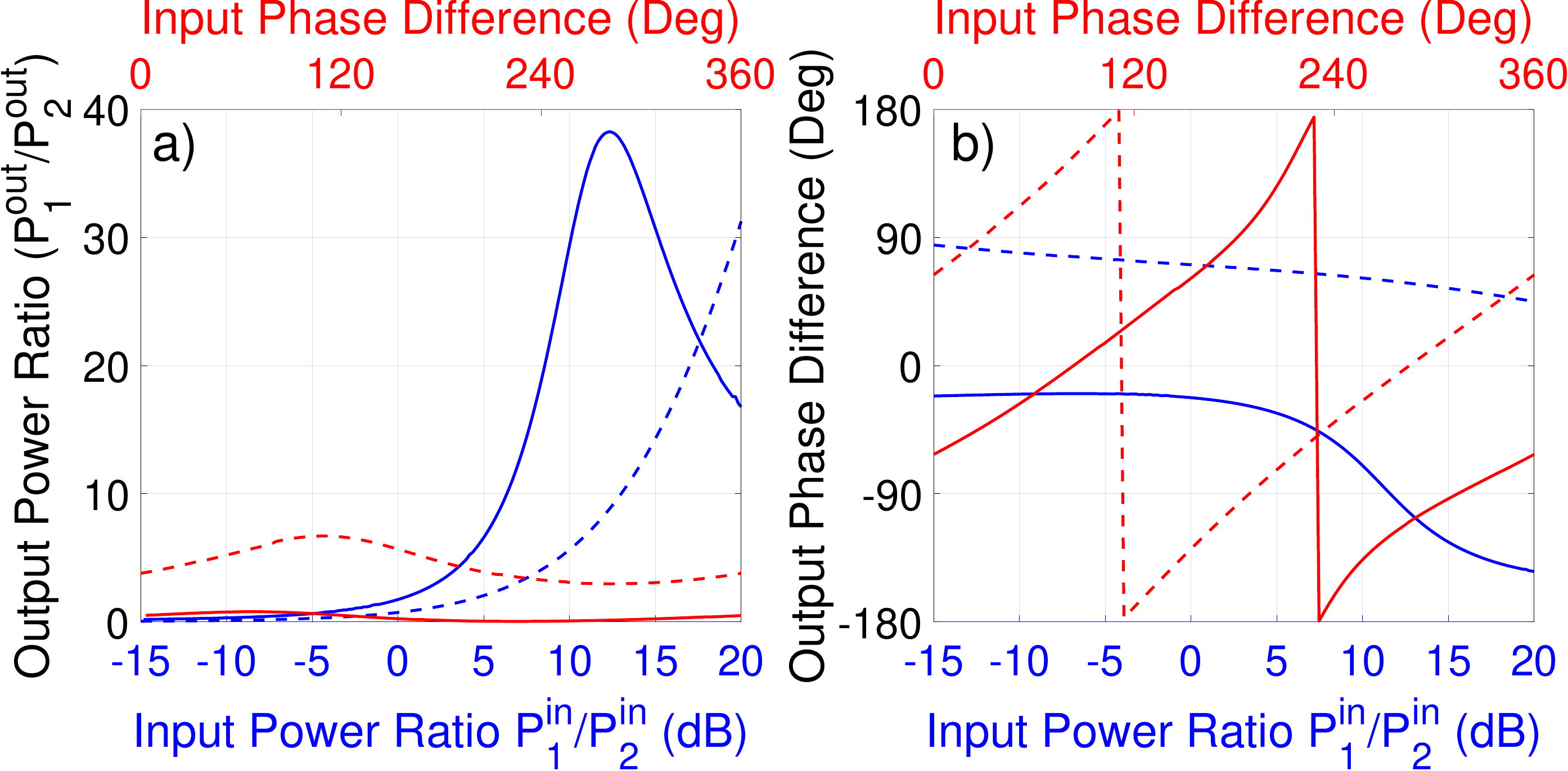}}
\caption{Experimental demonstration of arbitrary signals injected into a reciprocal microwave network far away from the CPA+EPD condition. a) Output power ratio vs input power ratio (lower axis, blue) and input phase difference (upper axis, red). The blue curves correspond to the input signals power ratio being swept, and the red curves correspond to the input signals relative phase being swept. The solid and dashed curves for each color denote two different arbitrary settings of the system parameters.  b) Output phase difference vs input power ratio (lower axis, blue) and input phase difference (upper axis, red). The blue and red curves (solid and dashed) have the same interpretation as in (a).  These results contrast dramatically with those shown in Fig. \ref{CPA_EP_Recip}.}
\label{Arbitrary_Injection}
\end{figure}

In Figure \ref{Arbitrary_Injection}, we show several representative examples of arbitrary signals injected into a complex graph not at a CPA or EPD condition to contrast with the phenomenon seen in Fig. \ref{CPA_EP_Recip}. For these arbitrary signals, we chose the settings of the phase shifters and frequency to be a random value, then did the same injection procedure specified above. As the system is not at any special condition, the output power ratio and phase difference between the two ports varies dramatically as the input signals change.

\clearpage
\newpage

\bibliography{Bibliography.bib}

\end{document}